\documentclass[namedreferences]{solarphysics}
\usepackage[optionalrh]{spr-sola-addons} 
\usepackage{graphicx}        
\usepackage{color}           
\usepackage{url}             
\usepackage{booktabs}             

\begin{document}
\begin{article}

\begin{opening}

\title{Validation and Benchmarking of a Practical Free Magnetic Energy and Relative Magnetic Helicity Budget Calculation in Solar Magnetic Structures}

\author{K.~\surname{Moraitis}$^1$\sep K.~\surname{Tziotziou}$^1$\sep M.~K.~\surname{Georgoulis}$^{1,2}$\sep V.~\surname{Archontis}$^3$       }
\runningauthor{Moraitis et al.}
\runningtitle{Validation of a practical calculation of free energy and relative helicity}

   \institute{$^{1}$ Research Center for Astronomy and Applied Mathematics
(RCAAM), Academy of Athens, 4 Soranou Efesiou Street, Athens, Greece, GR-11527
                     email: \url{kmorait@phys.uoa.gr} email: \url{kostas@noa.gr} email: \url{manolis.georgoulis@academyofathens.gr}\\
$^{2}$ Marie Curie Fellow\\
              $^{3}$ School of Mathematics and Statistics, St. Andrews University, St. Andrews, KY169SS, UK
                     email: \url{vasilis@mcs.st-and.ac.uk} \\
             }

\begin{abstract}
In earlier works we introduced and tested a nonlinear force-free (NLFF) method designed to self-consistently calculate the free magnetic energy and the relative magnetic helicity budgets of the corona of observed solar magnetic structures. The method requires, in principle, only a single, photospheric or low-chromospheric, vector magnetogram of a quiet-Sun patch or an active region and performs calculations in the absence of three-dimensional magnetic and  velocity-field information. In this work we strictly validate this method using three-dimensional coronal magnetic fields. Benchmarking employs both synthetic, three-dimensional magnetohydrodynamic simulations and nonlinear force-free field extrapolations of the active-region solar corona. We find that our time-efficient NLFF method provides budgets that differ from those of more demanding semi-analytical methods by a factor of $\sim 3$, at most. This difference is expected from the physical concept and the construction of the  method. Temporal correlations show more discrepancies that, however, are soundly improved for more complex, massive active regions, reaching correlation coefficients of the order of, or exceeding, 0.9. In conclusion, we argue that our NLFF method can be reliably used for a routine and fast calculation of free magnetic energy and relative magnetic helicity budgets in targeted parts of the solar magnetized corona. As explained here and in previous works, this is an asset that can lead to valuable insight into the physics and the triggering of solar eruptions.
\end{abstract}

\end{opening}

\section{Introduction}
     \label{S-Introduction}

Free magnetic energy and relative magnetic helicity in conjunction are nowadays believed \cite{tgr12,tgl13} to be two essential parameters for quantifying the eruptive capacity of solar active regions (ARs). Free magnetic energy quantifies the excess energy stored in a magnetic configuration with respect to its potential, current-free, energy state while magnetic helicity quantifies the twist, writhe and linkage of the magnetic field lines \cite[and references therein]{berg99}. Magnetic helicity, and more specifically relative (to the current-free field) magnetic helicity \cite{BergerF84}, is a signed (left- or right-handed) quantity. As such, while non-zero relative helicity always implies non-zero free magnetic energy, zero-helicity may imply non-zero free energy in the case of equal and opposite amounts of left- and right-handed helicities present in the magnetic configuration; current-free configurations result in both zero free magnetic energy and zero relative magnetic helicity. Since helicity dissipation is proportional to the inverse square of the magnetic Reynolds number \cite{freed93,berg99}, helicity is predominantly conserved during magnetic reconnection processes in the solar atmosphere, contrary to free magnetic energy which always dissipates during reconnection. If not transferred during reconnection either to larger scales via existing magnetic connections or to open magnetic field lines via interchange reconnection \cite{pari09,raou10} that afterwards release it to the heliosphere via unwinding motions, helicity can only be bodily removed from an AR in the form of coronal mass ejections \cite{low94,devore00}.

Simultaneous calculation of the instantaneous budgets of free magnetic energy and relative magnetic helicity in ARs can be achieved with a number of independent techniques, namely the a) time-integration of relative helicity and energy injection rates obtained via the Poynting theorem on the photospheric boundary (hereafter {\em flux-integration method}, e.g., \opencite{BergerF84}; \opencite{kusa02}), b) evaluation of the free energy and relative helicity formulas for the three-dimensional active-region corona (hereafter {\em volume-calculation method}, \opencite{BergerF84}; \opencite{fa85}), and c) translation of a single photospheric/chromospheric vector magnetogram to a single/ensemble of slender, force-free flux tubes and evaluation of its/their free energy and helicity with a linear (single tube) or a nonlinear (multiple tubes) force-free approximation (hereafter {\em LFF method}, \opencite{gbont07} or {\em NLFF method}, \opencite{geo12a}, respectively). Besides lacking a reference point at the start of the time-integration, the flux-integration method is susceptible to significant uncertainties stemming from the inference of a suitable photospheric flow-velocity field (e.g., \opencite{wels07}). On the other hand, the volume calculation method requires knowledge of the three-dimensional magnetic field configuration, typically acquired by extrapolations (e.g., \opencite{wieg04}), and its generating vector potential, including the corresponding current-free field and vector potential (e.g., \opencite{schm64}; \opencite{chae01}). However, nonlinear force-free field extrapolations are model-dependent and subject to uncertainties and ambiguities \cite{schrijver06,metc08,derosa09}, at the same time being computationally expensive and hence unattractive for routine application at high spatial resolution. The latter constraint is much more demanding in case nonlinear force-free field extrapolations are replaced by three-dimensional, data-driven magnetohydrodynamical (MHD) models
of observed ARs. The LFF method, introduced by \inlinecite{gbont07}, simultaneously calculates the relative magnetic helicity and the free magnetic energy of a magnetic structure represented by a single solar vector magnetogram. It is a physically self-consistent ``surface'' energy/helicity calculation, that does not rely on any prescribed flow velocity field or three-dimensional magnetic-field configuration. Its major weakness, nonetheless, is the adoption of the linear force-free field approximation which is, in principle, unrealistic for both the solar photosphere and the overlaying corona. The drawback of a LFF, constant-alpha parameter, has been tackled by the introduction of the NLLF method \cite{geo12a}, where the magnetic structure is translated into an ensemble of flux tubes characterized by different fluxes and $\alpha$-parameters. This has led to a novel, self-consistent and unique (per a given set of flux tubes) surface-calculation of the free magnetic energy and the relative magnetic helicity. Uniqueness is guaranteed by an iterative simulated annealing method that optimizes the flux-tube connectivity following a set of
rules reflected on the minimization of a given functional. Simulated annealing is known to converge on the exact absolute minimum of this, or any given, functional if implemented for a sufficiently high number of iterations (e.g., \opencite{metrop53}; \opencite{kirkp83}). Given the lack of knowledge of the three-dimensional coronal magnetic structure, the NLFF method calculates a lower limit for the free magnetic energy and a correspondingly constrained relative magnetic helicity, as explained in \inlinecite{geo12a}. The method has already given interesting results, such as the energy-helicity diagram of solar ARs \cite{tgr12} and the physical interpretation of eruptive ARs, such as NOAA AR 11158 \cite{tgl13}.

Meaningful results aside, the NLFF method has been validated only in a rather rudimentary way, namely by correlating results with respective volume-calculation results from observed/extrapolated ARs (see Section 3 of \opencite{geo12a}). A rigorous validation and benchmarking of the NLFF method, that aims to calculate the inherently three-dimensional magnetic energy and helicity budgets requiring neither a three-dimensional coronal magnetic field nor a two-dimensional lower-boundary velocity field, is
the central task of this study. To validate our method we use synthetic, MHD-simulated magnetic fields of an eruptive and a non-eruptive AR (e.g., \opencite{arch04}), where the overlaying three-dimensional ``coronal'' field is accurately known. This can provide comparisons between the NLFF and volume-calculation methods without the need for model-dependent extrapolations. In addition, we use observed vector magnetograms of an eruptive and a non-eruptive solar AR, where the comparison between the NLFF and the volume-calculation methods must incorporate the results of less reliable, three-dimensional field extrapolations. As mentioned above, this test is robust because the volume-calculation method relies on semi-analytical, generally accepted formulas for energy and helicity that have all necessary information incorporated. The test will allow us to quantitatively assess the loss of information, in terms of energy and helicity budgets, imposed by the use of the simplifying, but readily applicable NLFF method.

Section~\ref{S-features} discusses our approach to apply the volume-calculation and NLFF surface methods. Section~\ref{S-general} describes the synthetic and observed data, while Section~\ref{S-results} presents the results and their implications. Section~\ref{S-discuss}, summarizes the study and outlines our conclusions.

\section{Method description}
      \label{S-features}

In this section we describe the two methods we use to calculate the instantaneous free magnetic energy and relative magnetic helicity budgets of bounded, three-dimensional magnetic structures.

\subsection{Numerical NLFF method}
  \label{S-equations}

The NLFF method of \inlinecite{geo12a} combines the LFF method of \inlinecite{gbont07} with the properties of mutual helicity as described by \inlinecite{dem06}. The method uses a single photospheric or chromospheric vector magnetogram to derive a unique magnetic-connectivity matrix that contains the flux committed to connections between positive- and negative-polarity flux partitions. This connectivity matrix, since the three-dimensional magnetic configuration is unknown, is derived by means of a simulated annealing method introduced by \inlinecite{geo:rus}, a method that guarantees {\em both} connections between opposite-polarity flux partitions and global minimization of the corresponding connection lengths. This implementation favors complex active regions with intense, highly sheared magnetic polarity inversion lines. Inference of the connectivity matrix could also be based on line tracing of individual magnetic field lines derived by a nonlinear force-free field extrapolation method or an MHD model. However, contrary to the aforementioned annealing approach, line-tracing results would not be unique as they are a) model-dependent and b) several closed field lines extend beyond the boundaries of the finite volume, so line-tracing would erroneously treat these lines as open.

All non-zero flux elements of the connectivity matrix define a collection of $N$ magnetic connections. These connections are treated as slender force-free flux tubes with known footpoints, flux contents, and variable force-free parameters $\alpha$. Assuming that there is no winding of flux tubes around each other's axes, that is equivalent to assuming that the unknown Gauss linking number is set to zero, \inlinecite{geo12a} expressed the free magnetic energy $E_\mathrm{c}$ as the sum of a self term $E_{\mathrm{c}_{\rm self}}$, that describes the internal twist and writhe of each flux tube, and a mutual term $E_{\mathrm{c}_{\rm mut}}$, that describes interactions between different flux tubes. For a zero Gauss linking number, the free energy represents a lower but realistic limit \cite{geo12a,tgr12}, and is given by
\begin{eqnarray}
E_\mathrm{c}  & = & E_{\mathrm{c}_{\rm self}} + E_{\mathrm{c}_{\rm mut}} \nonumber \\
 & = & A \lambda^2 \sum _{l=1}^N \alpha _l^2
\Phi_l^{2 \delta} +
      \frac{1}{8 \pi} \sum _{l=1}^N \sum _{m=1, l \ne m}^N
           \alpha _l \mathcal{L}_{lm}^{\rm arch} \Phi_l \Phi_m\;\;,
\label{Ec_fin}
\end{eqnarray}
where $A$ and $\delta$ are known fitting constants, $\lambda$ is the pixel size of the magnetogram, $\Phi_l$ and $\alpha_l$ are the respective unsigned flux and
force-free parameters of flux tube $l$, and $\mathcal{L}_{lm}^{\rm arch}$ is the mutual-helicity factor. This is a factor describing the ``interaction" of two arch-like flux tubes, first introduced by \inlinecite{dem06} and further analyzed by \inlinecite{geo12a} for all possible cases.

The respective relative magnetic helicity $H$ for a collection of $N$ slender flux tubes is the sum of a self $H_{\rm self}$ and a mutual $H_{\rm mut}$ term \cite{geo12a}:
\begin{eqnarray}
H  & = & H_{\rm self} + H_{\rm mut} \nonumber \\
&  = & 8 \pi \lambda^2 A
\sum _{l=1}^N \alpha _l \Phi_l ^{2 \delta} +
      \sum _{l=1}^N \sum _{m=1,l \ne m}^N \mathcal{L}_{lm}^{\rm arch} \Phi_l
      \Phi_m\;\;.
\label{Hm_fin}
\end{eqnarray}

Derivation of uncertainties for all terms of the free magnetic energy and the relative magnetic helicity is fully described in \inlinecite{geo12a}.

\subsection{Semi-analytical free magnetic energy and relative magnetic helicity budget calculation}
  \label{S-simple-equations}

The numerical NLFF method of Section~\ref{S-equations} will be tested against well-known analytical expressions for the free magnetic energy and relative magnetic helicity. There are two equivalent ways to express the excess, free energy $E_\mathrm{c}$ of a three-dimensional magnetic field $\mathbf{B}$ occupying a finite, bounded volume $\mathcal{V}$ with boundary $\partial \mathcal{V}$ \footnote{The same methodology can be generalized for a semi-infinite (i.e., partly bounded) volume, where the field becomes zero at infinity.}, relative to the current-free (potential) magnetic field $\mathbf{B}_\mathrm{p}$: first, by considering the energy difference between the two fields $\mathbf{B}$ and $\mathbf{B}_\mathrm{p}$, namely
\begin{equation}
E_\mathrm{c}=E_\mathrm{t}-E_\mathrm{p}=\frac{1}{8\pi}\int_\mathcal{V} \mathrm{d}\mathcal{V}\,\mathbf{B}^2-\frac{1}{8\pi}\int_\mathcal{V} \mathrm{d}\mathcal{V}\,\mathbf{B}_\mathrm{p}^2
\label{ecdef}
\end{equation}
and, second, by considering the volume energy of the vector difference between these two fields (i.e., the current-carrying or non-potential vector field), namely
\begin{equation}
E'_\mathrm{c}=\frac{1}{8\pi}\int_\mathcal{V} \mathrm{d}\mathcal{V}\,(\mathbf{B}-\mathbf{B}_\mathrm{p})^2.
\label{ecdef2}
\end{equation}
The difference between Equation~(\ref{ecdef}) and Equation~(\ref{ecdef2}) relates to the volume integral of the scalar product $\mathbf{B}_\mathrm{c}\cdot\mathbf{B}_\mathrm{p}$, where $\mathbf{B}_\mathrm{c}=\mathbf{B}-\mathbf{B}_\mathrm{p}$ is the current-carrying magnetic field vector. Since the potential field can be written as $\mathbf{B}_\mathrm{p}=-\nabla \varphi$ with $\varphi$ a scalar function, it can easily be shown that
\begin{equation}
E_\mathrm{c}-E'_\mathrm{c}=-\frac{1}{4\pi}\int_{\partial\mathcal{V}}\varphi\mathbf{B}_\mathrm{c}\cdot \mathrm{d}\mathbf{S}+\frac{1}{4\pi}\int \mathrm{d}\mathcal{V}\varphi(\nabla\cdot \mathbf{B}_\mathrm{c}),
\label{ecdiff}
\end{equation} 
with $\mathrm{d} \mathbf{S}$ being the infinitesimal surface area on $\partial \mathcal{V}$. As we will see in Appendix~\ref{S-simple-equations-errors}, this will essentially provide an uncertainty for the free-energy calculation in the volume-calculation method.

The relative (to the reference field $\mathbf{B}_\mathrm{p}$) magnetic helicity of the field $\mathbf{B}$ in the given volume $\mathcal{V}$ is provided by the well-known \inlinecite{fa85} relation
\begin{equation}
H=\int_\mathcal{V} \mathrm{d}\mathcal{V}\,(\mathbf{A}+\mathbf{A}_\mathrm{p})\cdot(\mathbf{B}-\mathbf{B}_\mathrm{p})
\label{tothel}
\end{equation}
where $\mathbf{A}$, $\mathbf{A}_\mathrm{p}$ are the generating vector potentials of the fields $\mathbf{B}$ and $\mathbf{B}_\mathrm{p}$, respectively. This quantity is gauge-independent even for gauge-dependent $\mathbf{A}$, $\mathbf{A}_\mathrm{p}$ as long as $\mathbf{B}_\mathrm{p}$ has the same normal components as $\mathbf{B}$ on the boundaries of the volume, and both fields are divergence-free. Relative helicity can be split into two parts \cite{berg99}: self helicity, owing to the twist and writhe of individual flux tubes, given by
\begin{equation}
H_\mathrm{self}=\int_\mathcal{V} \mathrm{d}\mathcal{V}\,(\mathbf{A}-\mathbf{A}_\mathrm{p})\cdot(\mathbf{B}-\mathbf{B}_\mathrm{p})
\label{hselfdef}
\end{equation}
and mutual helicity, representing the interaction between pairs of flux tubes, given by
\begin{equation}
H_\mathrm{mut}=2\int_\mathcal{V} \mathrm{d}\mathcal{V}\,\mathbf{A}_\mathrm{p}\cdot(\mathbf{B}-\mathbf{B}_\mathrm{p}).
\label{hmutdef}
\end{equation}
In the following we calculate the various energy and helicity budgets directly from Equations~(\ref{ecdef})-(\ref{hmutdef}) using as input a three-dimensional magnetic field $\mathbf{B}$ that may be provided by nonlinear force-free field extrapolations or by MHD simulations. In order to do this we need to know three additional vector fields, namely the two vector potentials $\mathbf{A}$, $\mathbf{A}_\mathrm{p}$, and the potential field $\mathbf{B}_\mathrm{p}$. The calculation of these fields is achieved as explained below.

\subsubsection{Calculation of the potential field}
\label{S-potfld}

In calculating the potential field we choose a numerical procedure that takes into account all boundaries of the given finite volume, rather than using the \inlinecite{schm64} method, which is valid for the semi-infinite space above a lower boundary. We denote the rectangular volume of interest as $\mathcal{V}=(x_1,x_2)\times(y_1,y_2)\times(z_1,z_2)$. The potential field satisfies $\nabla\times\mathbf{B}_\mathrm{p}=0$ in $\mathcal{V}$, so that the electric current density vanishes. This implies that the potential field can be expressed as $\mathbf{B}_\mathrm{p}=-\nabla \varphi$, where $\varphi$ is a scalar potential. The additional zero-divergence condition for the potential field translates into the scalar potential being a solution of Laplace's equation $\nabla^2\varphi=0$ in $\mathcal{V}$. The requirement that $\mathbf{B}$ and $\mathbf{B}_\mathrm{p}$ have the same normal components along the boundaries of the volume then leads to the Neumann boundary conditions for $\varphi$
\begin{equation}
\left.\frac{\partial\varphi}{\partial\hat{n}}\right|_{\partial \mathcal{V}}=-\left.\hat{n}\cdot\mathbf{B}\right|_{\partial \mathcal{V}}.
\end{equation}
We solve Laplace's equation numerically using a standard FORTRAN routine included in the FISHPACK library \cite{ss79}. We note however that the solution of Laplace's equation under Neumann boundary conditions is guaranteed to exist (up to an additive constant) only for valid field solutions, i.e. for fields satisfying $\int_{\partial \mathcal{V}}\mathbf{B}\cdot\mathrm{d}\mathbf{S}=0$. This is equivalent to reiterating the validity of the divergence-free condition for the magnetic field vector. If this condition is compromised, which could be the case in imperfect, numerical or optimization-based extrapolated fields, then results should be treated with extreme caution. 

The boundary conditions imposed on $\varphi$ guarantee that the current-carrying part of the magnetic field is enclosed in the given volume, i.e., $\left. \hat{n}\cdot\mathbf{B}_\mathrm{c}\right|_{\partial \mathcal{V}}=0$. It then follows from Equation~(\ref{ecdiff}) that the difference between $E_\mathrm{c}$ and $E'_\mathrm{c}$ is a measure of the errors in the solenoidal property of $\mathbf{B}$, mainly, and $\mathbf{B}_\mathrm{p}$, secondarily.

\subsubsection{Calculation of vector potentials}

With the original and potential fields at hand, we now proceed to calculate the corresponding vector potentials. To this we follow the method proposed by \inlinecite{val12}: we first choose the gauge $\hat{z}\cdot\mathbf{A}=0$ throughout $\mathcal{V}$, so that the $x$ and $y$ components of $\mathbf{B}=\nabla\times\mathbf{A}$ can be integrated in the interval $(z_1,z)$ to
\begin{equation}
\mathbf{A}=\mathbf{A}_0-\hat{z}\times\int_{z_1}^z\mathrm{d}z'\,\mathbf{B}(x,y,z').
\label{vpot}
\end{equation}
The integration vector $\mathbf{A}_0=\mathbf{A}(x,y,z=z_1)=(A_{0x},A_{0y},0)$ satisfies the $z$ component of $\mathbf{B}=\nabla\times\mathbf{A}$, namely
\begin{equation}
\frac{\partial A_{0y}}{\partial x} -\frac{\partial A_{0x}}{\partial y} =B_z(x,y,z=z_1).
\label{vpot2}
\end{equation}
The simplest solution to Equation~(\ref{vpot2}) is given by
\begin{eqnarray}
A_{0x}=-\frac{1}{2}\int_{y_1}^y\mathrm{d}y'\,B_z(x,y',z=z_1)\label{vpot3}\\
A_{0y}=\frac{1}{2}\int_{x_1}^x\mathrm{d}x'\,B_z(x',y,z=z_1)\label{vpot4}.
\end{eqnarray}
In deriving Equation~(\ref{vpot2}), a divergence-free field is assumed. Again, any numerical compromise of this condition seriously impacts 	the validity of these formulas.

An alternative solution for the vector potential can be obtained if we integrate in the interval $(z,z_2)$. In this case, the result is	
\begin{equation}
\mathbf{A}=\mathbf{\tilde{A}}_0+\hat{z}\times\int_{z}^{z_2}\mathrm{d}z'\,\mathbf{B}(x,y,z')
\end{equation}
where $\mathbf{\tilde{A}}_0=\mathbf{A}(x,y,z=z_2)$ satisfies again Equation~(\ref{vpot2}), but with $B_z$ calculated at $z=z_2$. In the following, we employ both formulations and denote which one we use. For the calculation of the vector potential for the potential field, $\mathbf{A}_\mathrm{p}$, we follow the same procedure and note that $\mathbf{A}_{\rm p0}=\mathbf{A}_0$ (or $\mathbf{\tilde{A}}_{\rm p0}=\mathbf{\tilde{A}}_0$ for the alternative solution), since $\mathbf{B}$, $\mathbf{B}_\mathrm{p}$ share the same normal components at $z=z_1$ (or at $z=z_2$).

All integrations are done with a modified Simpson's rule with error estimate of order $1/N^4$ \cite{nrs} where $N\geq 3$ is the number of points in the integration, while for $N=2$ the trapezoidal rule is used. Also, all differentiations are done using the appropriate (centered, forward or backward) second-order numerical derivative. The use of specific integration and differentiation rules contrasts the efforts of \inlinecite{val12} to achieve numerical reversibility of differentiation and integration, as these two procedures cannot be numerically reversible. Nonetheless, the results obtained with our prescription and the one of \inlinecite{val12} differ insignificantly.

Estimation of the errors included in the semi-analytical method are given in Appendix~\ref{S-simple-equations-errors}, while an analysis of its performance is given in Appendix~\ref{S-apdx}.

\section{Data description}
      \label{S-general}

\subsection{Numerical MHD experiments}
  \label{S-text}

We have performed two numerical experiments of magnetic flux emergence of a twisted magnetic flux tube from the solar interior into the solar corona. 
The emergence of the field at the photosphere leads to the formation of a small AR (e.g. \opencite{arch04}). In both experiments the simulation box has dimensions $65\times65\times65$~Mm. In the first experiment, the simulated AR is non-eruptive and is modeled for $\sim 9.5$~h of real solar time, while the second simulation is for an eruptive AR and covers its evolution up to $\sim 4.5$~h of real time (for an analytic description of the eruptive AR see \opencite{aht14}). By ``eruptive'' AR we imply an active-region case where part of the magnetic structure is ejected beyond the modeled volume at least once in the course of the simulation. In our ``non-eruptive'' active-region case the magnetic structure remains confined within the modeled volume for the duration of the simulation.

To perform the experiments we solve the three-dimensional time-dependent and resistive MHD equations in Cartesian geometry:
\begin{eqnarray} \label{ConsMass}
&\frac{\partial\rho}{\partial t} + \nabla \cdot (\rho {\bf v}) = 0\\
&\frac{\partial (\rho {\bf v})}{\partial t} = -\nabla\cdot(\rho{\bf
v}{\bf v})+(\nabla \times {\bf B})\times{\bf B}-\nabla {P}+\rho{\bf g}+\nabla\cdot{\bf S}\\
&\frac{\partial (\rho \epsilon)}{\partial t} = -\nabla
\cdot(\rho\epsilon{\bf v}) -P\nabla\cdot{\bf v} + Q_{\mathrm{Joule}} +
Q_{\mathrm{visc}}\\
&\frac{\partial {\bf B}}{\partial t} = \nabla \times \bigl( {\bf v}
\times {\bf B} \bigr) + \eta \nabla^2{\bf B}
\end{eqnarray}
with specific energy density
\begin{equation} \label{energy_density}
\epsilon = \frac{P}{(\gamma-1)\rho}.
\end{equation}

${\bf B}$, $\rho$, $P$ and ${\bf v}$ denote the magnetic field vector, the density, the gas pressure and the velocity vector, respectively. Gravity is included, with ${\bf g} = - g \hat{\bf{z}}$ being the gravitational acceleration and $g=274 \,\mathrm{m}\,\mathrm{s}^{-2}$. For the explicit dimensionless resistivity we use a constant value of $\eta=10^{-3}$. The medium is assumed to be a perfect gas with a ratio of specific heats $\gamma=5/3$. Viscous and Ohmic heating are considered through the viscosity and Joule dissipation terms, $Q_{\mathrm{visc}}$ and $Q_{\mathrm{Joule}}$ respectively, while $\mathbf{S}$ represents the viscous stress tensor. The above equations are numerically solved by using the Lare3d code \cite{arber01}. For more details on the equations and the numerical setup of similar flux emergence experiments see also the work by \inlinecite{arber07} and \inlinecite{arc13}, respectively.

The above equations are written in dimensionless form. For the conversion into dimensional variables, we use the following units: density $\rho=1.67\times10^{-7}\,\mathrm{g}\,\mathrm{cm}^{-3}$, temperature $T=5100\,\mathrm{K}$ and pressure $P=7.16\times10^{4}\,\mathrm{erg}\,\mathrm{cm}^{-3}$. For the length, we use $\lambda=180\,\mathrm{km}$ and for the magnetic field strength $B=300\,\mathrm{G}$. Using the above units we obtain the velocity and time units, $v=2.1\,\mathrm{km}\,\mathrm{s}^{-1}$ and $t=85.7\,\mathrm{s}$, respectively.

\begin{figure}
\includegraphics[width=\textwidth,clip]{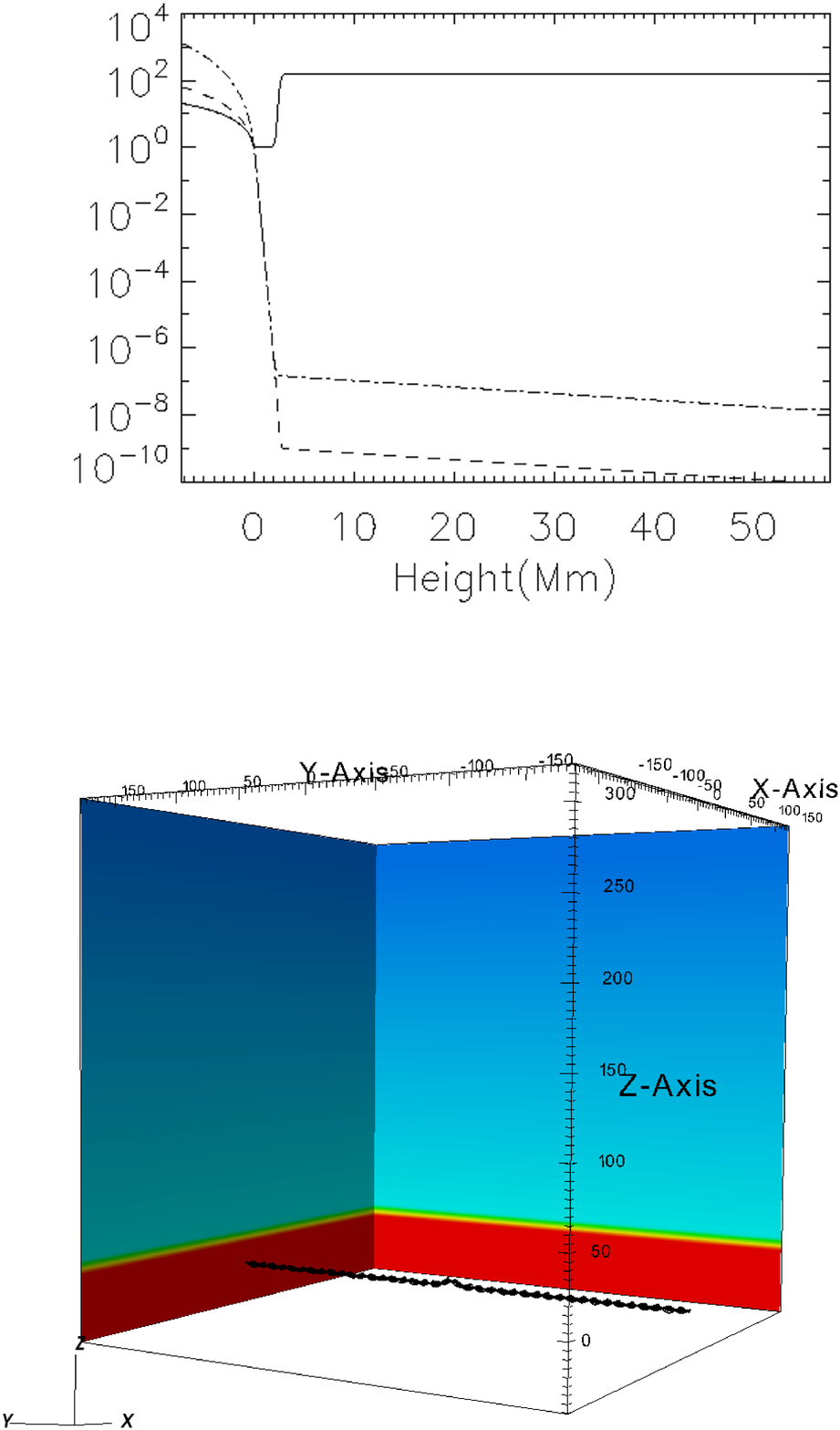}
\caption{{\bf Top:} The (dimensionless) stratification of the initial atmosphere: temperature (solid), pressure (dot-dashed) and density (dashed).
{\bf Bottom:} The twisted magnetic flux tube is visualized by a set of fieldlines, which are traced from the sub-photospheric footpoints of the 
tube ($t\approx 9.5\,\mathrm{min}$). At this stage of the evolution, the tube has started to emerge towards the solar surface. The vertical slices illustrate the density stratification across the background atmosphere (red:high density, blue/cyan:low density).}
\label{fig1_vas}
\end{figure}

The initial conditions in the MHD model consist of the background hydrostatic atmosphere and a horizontal twisted magnetic flux tube below the photosphere.
Figure \ref{fig1_vas} (top panel) shows the initial distribution of the temperature ($T$), density ($\rho$) and gas pressure ($P$) as functions of height. Hydrostatic equilibrium is assumed for the initial atmosphere. The sub-photospheric layer is represented by an adiabatically stratified layer in the range ($-7.2\,\mathrm{Mm} \leq z < 0\,\mathrm{Mm}$).
The photosphere/chromosphere is represented by a layer at $0\,\mathrm{Mm}\leq z < 2.3\,\mathrm{Mm}$, which is isothermal at the beginning and then the temperature increases with height, up to $\approx 4\times10^{4}\,\mathrm{K}$.
The layer above, at $2.3\,\mathrm{Mm}\leq z \leq 3.1\,\mathrm{Mm}$, is mimicking the transition region. The uppermost layer ($3.1\,\mathrm{Mm} < z \leq 58\,\mathrm{Mm}$) is an isothermal layer ($\approx 1\,\mathrm{MK}$) that represents the corona.

For the initial magnetic field, we considered a horizontal cylindrical magnetic flux tube (bottom panel, Figure \ref{fig1_vas}), which is located $2.1\,\mathrm{Mm}$ below the photosphere. The flux tube is oriented with its axis along the positive $y$-direction. The transverse direction is $x$ and the vertical direction is $z$. The axial field, $B_\mathrm{y}$ is defined by
\begin{equation}
B_y=B_0\,\mathrm{exp}(-r^2/R_\mathrm{t}^2)
\label{by}
\end{equation}
and the azimuthal field by
\begin{equation}
B_\phi=\alpha\,r\,B_y,
\label{bphi}
\end{equation}
where $R_\mathrm{t}$ is a measure of the radius of the tube, $r$ is the radial distance from the tube axis, and $B_0$ is the magnetic field strength on the axis. In the following, we chose $R_\mathrm{t}=2.5$ (i.e. $450\,\mathrm{km}$). To initiate the rising motion of the tube, we apply the same density distribution as in previous studies (e.g. \opencite{arc12}) that makes the middle part of the tube underdense and, hence, buoyant.

For the non-eruptive case, we use the buoyant part of the tube to have an approximate length of $L=10$ (i.e. $1.8\,\mathrm{Mm}$). We choose a uniform twist with $\alpha=5.5\times10^{-4}\,\mathrm{km}^{-1}$, which
implies that we study the evolution of a {\it weakly twisted} magnetic flux system (e.g. \opencite{arc13}). For the initial field strength of the subphotospheric flux tube, we use $B_0=2500\,\mathrm{G}$. 

For the eruptive case, we use a stronger field strength ($B_0=3150\,\mathrm{G}$), a higher twist ($\alpha=2.2\times10^{-3}\,\mathrm{km}^{-1}$) for which the flux tube is initially stable to the kink instability, and a smaller $L$ (i.e. $L=0.9\,\mathrm{Mm}$), which leads to more effective draining of heavy plasma from the apex of the tube during the emergence from the solar interior.

The numerical grid has 416 nodes in all directions with periodic boundary conditions in $y$. Open boundary conditions have been implemented along $x$
and at the top of the numerical domain, allowing outflow of plasma and magnetic field. The bottom boundary is a non-penetrating, perfectly conducting wall.

\begin{figure}
\begin{center}
\begin{minipage}[t]{0.48\textwidth}
\vspace{0pt}
\centering
\includegraphics[width=0.8\textwidth,height=0.8\textwidth]{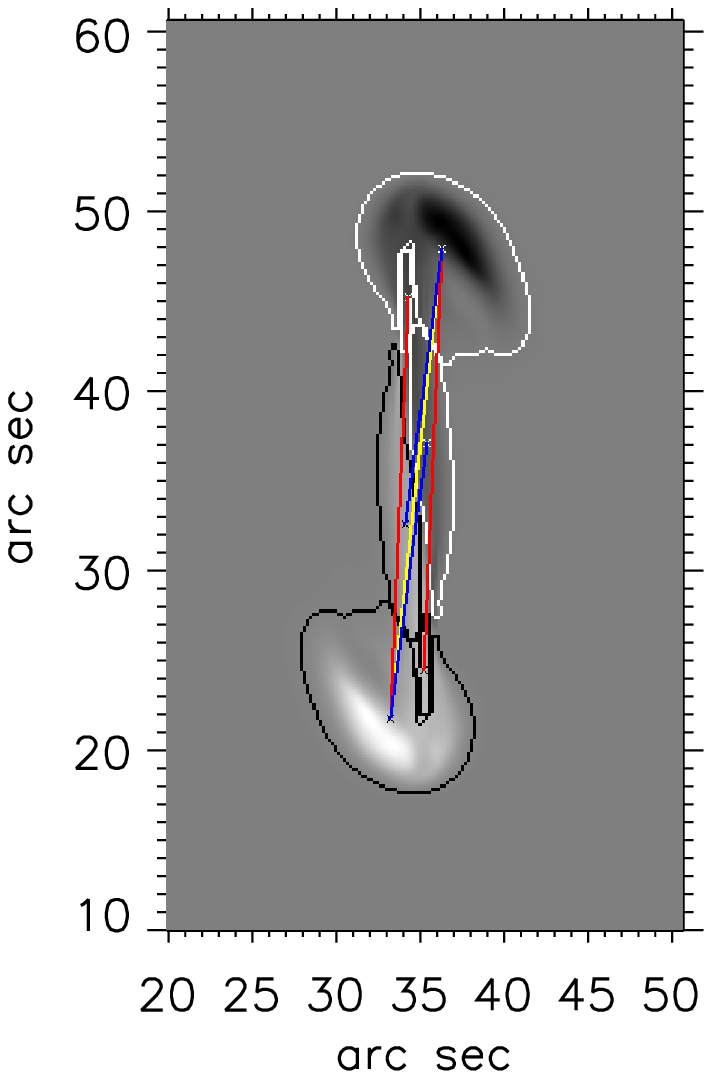}
\end{minipage}%
\begin{minipage}[t]{0.48\textwidth}
\vspace{0pt}
\centering
\includegraphics[width=\textwidth,height=0.8\textwidth]{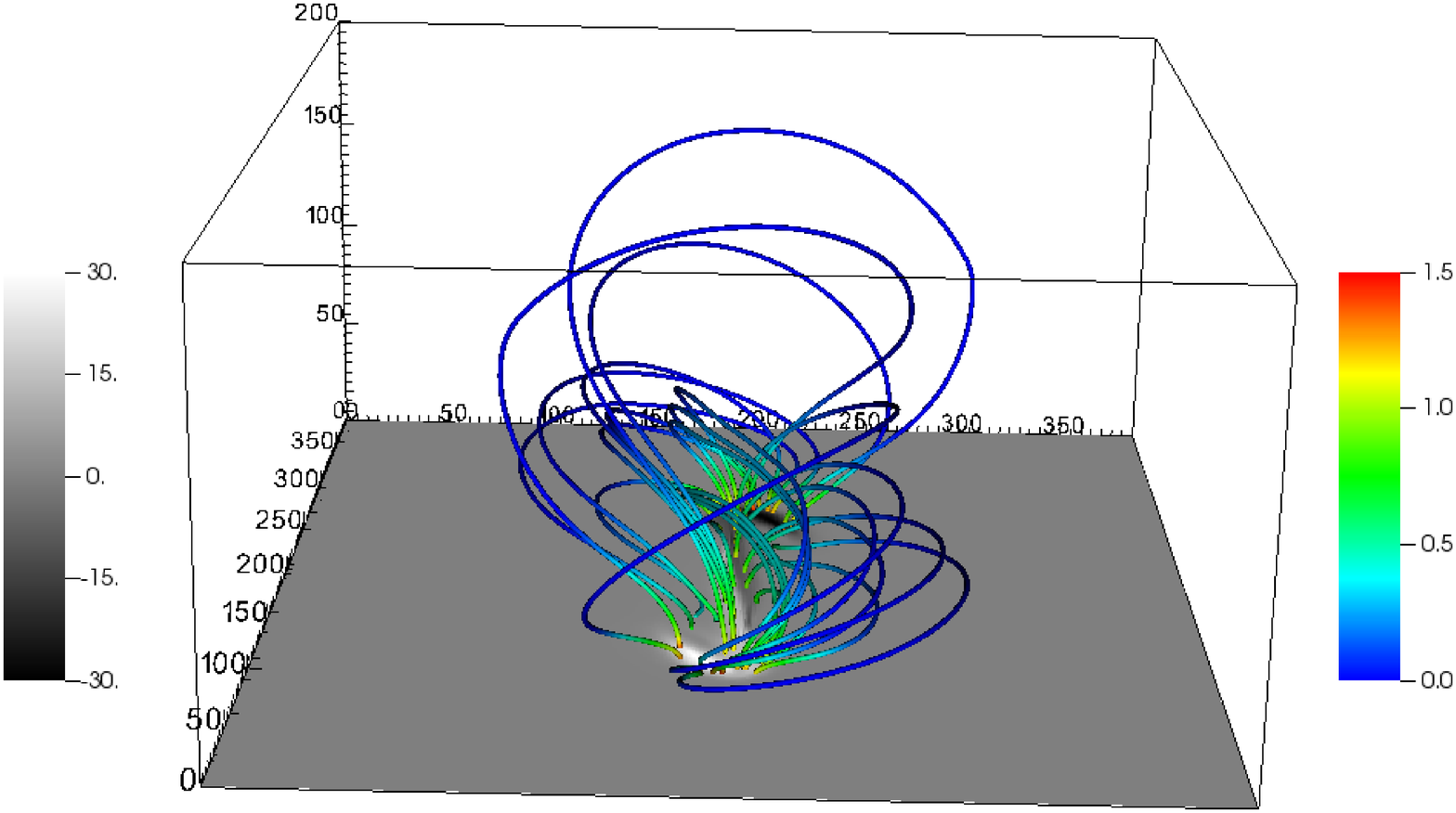}
\end{minipage}
\caption{{\bf Left panel:} Inferred magnetic connectivity of the synthetic non-eruptive AR at time $t\sim5.9$~h (see Section~\ref{S-text}), showing the vertical magnetic field component in grayscale with the contours bounding the identified magnetic partitions. Line segments represent flux-tube connections, identified by the magnetic connectivity matrix, closing within the field-of-view and connecting the flux-weighted centroids of the respective pair of partitions. Red, blue, yellow, and cyan segments denote magnetic flux contents within the ranges $<5\times10^{17}$ Mx, [5$\times10^{17}$, 10$^{18}$] Mx, [10$^{18}$, 5$\times 10^{18}$] Mx, and $>5\times 10^{18}$ Mx, respectively. {\bf Right panel:} Three-dimensional modeled structure of the magnetic field, with magnetic field lines coloured according to the logarithm of the magnetic field strength. Spatial scales are in units of pixels.}\label{figdata1}
\end{center}
\end{figure}

\begin{figure}
\begin{center}
\begin{minipage}[t]{0.48\textwidth}
\vspace{0pt}
\centering
\includegraphics[width=0.8\textwidth,height=0.8\textwidth]{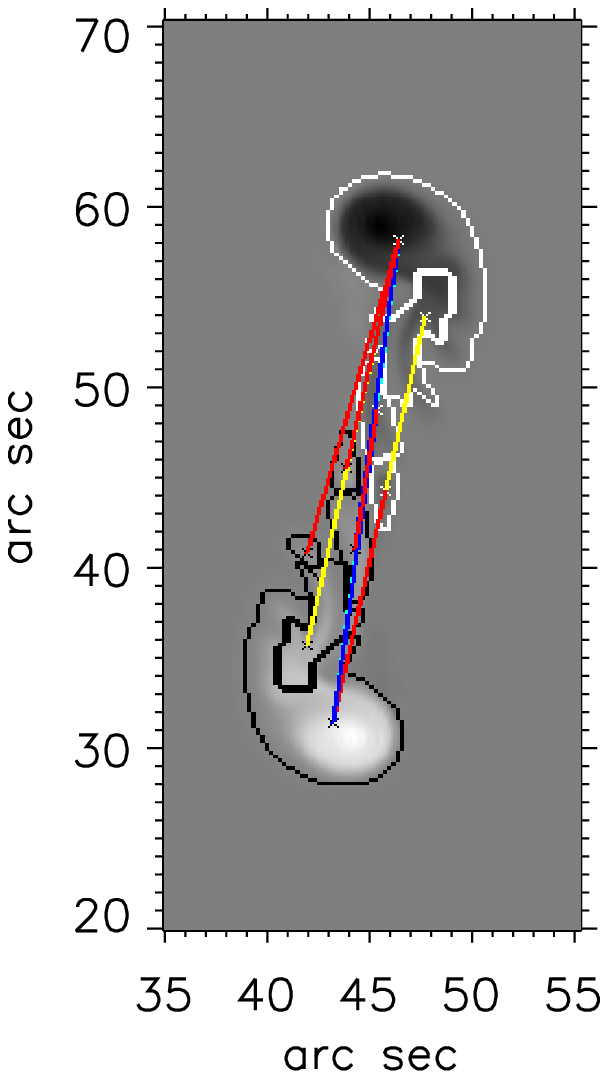}
\end{minipage}%
\begin{minipage}[t]{0.48\textwidth}
\vspace{0pt}
\centering
\includegraphics[width=\textwidth,height=0.8\textwidth]{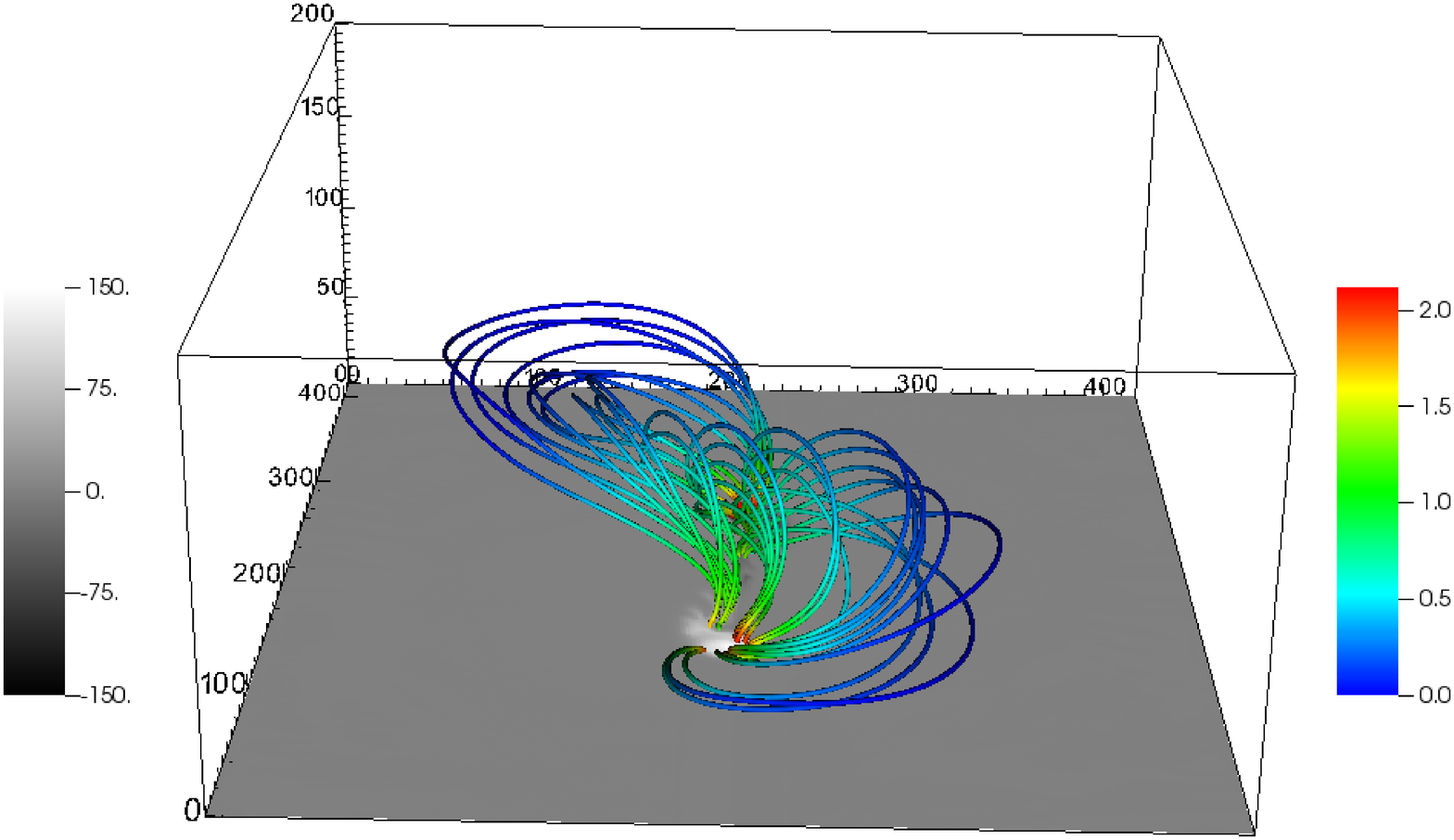}
\end{minipage}
\caption{Same as Figure~\ref{figdata1}, but for the synthetic eruptive AR at time $t\sim3.8$~h.}\label{figdata2}
\end{center}
\end{figure}

In Figures~\ref{figdata1}, \ref{figdata2} we show a snapshot of the three-dimensional structure of the magnetic field along with the NLFF-inferred photosperic magnetic connectivity, for the synthetic non-eruptive and eruptive ARs respectively.

\subsection{Observations and nonlinear force-free field extrapolations of NOAA ARs 11072 and 11158}
\label{S-obs}

For our analysis we use vector magnetic field observations of two observed solar ARs, the non-eruptive NOAA AR 11072 and the eruptive NOAA AR 11158, acquired with the Helioseismic and Magnetic Imager (HMI; \opencite{sche12}) onboard the Solar Dynamics Observatory (SDO; \opencite{pes12}). HMI samples the Fe~I 617.3~nm photospheric line through a 76 m\AA\ filter at six wavelength positions along the line (covering a range of $\lambda_{\rm 0}\pm$17.5~pm), with two CCD cameras, providing full disk (4096$\times$4096) filtergrams with a 0.5 arcsec pixel size. The first CCD camera is used for obtaining Dopplergrams and line-of-sight magnetograms at a cadence of 45~s from filtergrams recorded at the aforementioned six wavelength positions in two polarization states, while the second camera acquires six polarization states every 135~s, used to compute the four Stokes parameters ({\em I, Q, U,} and {\em V}). In order to enhance the signal-to-noise ratio, averaged 720-second filtergrams are used for the derivation of the Stokes parameters. The latter are necessary for deriving the vector magnetic field with a Milne-Eddington-based inversion approach \cite{borr11}.

Cutout vector magnetogram data were made available by SDO's Joint Science Operations Center. We use 22 vector magnetograms of NOAA AR 11072 covering a 5.25-day period (2010 May 20-25) with a 6-hour cadence. NOAA AR 11072 is an AR that exhibited no significant flaring activity (only six small B-class flares, the largest being a B6.5), despite continuously evolving via flux emergence. The photospheric area covered by these magnetograms is 512$\times$512 pixels, or 256$\times$256 arcsec on the image (observer's) plane. These magnetograms were rebinned to 128$\times$128-pixel magnetograms, covering the same photospheric area (pixel size of $\sim2$ arcsec) to enable performing and comparing of results with nonlinear force-free magnetic field extrapolations. Performing these extrapolations in the original magnetogram resolution would be very demanding computationally. The vector magnetograms were treated for the azimuthal 180$^{\rm o}$ ambiguity using the non-potential field calculation of \inlinecite{geo05}, as revised in \inlinecite{metc06}. For the analysis we use the heliographic components of the magnetic field vector on the heliographic plane, derived with the de-projection equations of \inlinecite{gary90}. As typical single-value uncertainties for the line-of-sight and transverse field components and for the azimuth angle ($\delta B_l$, $\delta B_{tr}$ and $\delta \phi$ respectively), we use $\delta B_l\sim5$ G, $\delta B_{tr}\sim50$ G and $\delta\phi\sim0^{\rm o}$. These uncertainties are used in the computation of the error for the NLFF method, as mentioned in Section~\ref{S-equations} and described in \inlinecite{geo12a}.

\begin{figure}
\begin{center}
\begin{minipage}[t]{0.48\textwidth}
\vspace{0pt}
\centering
\includegraphics[width=\textwidth]{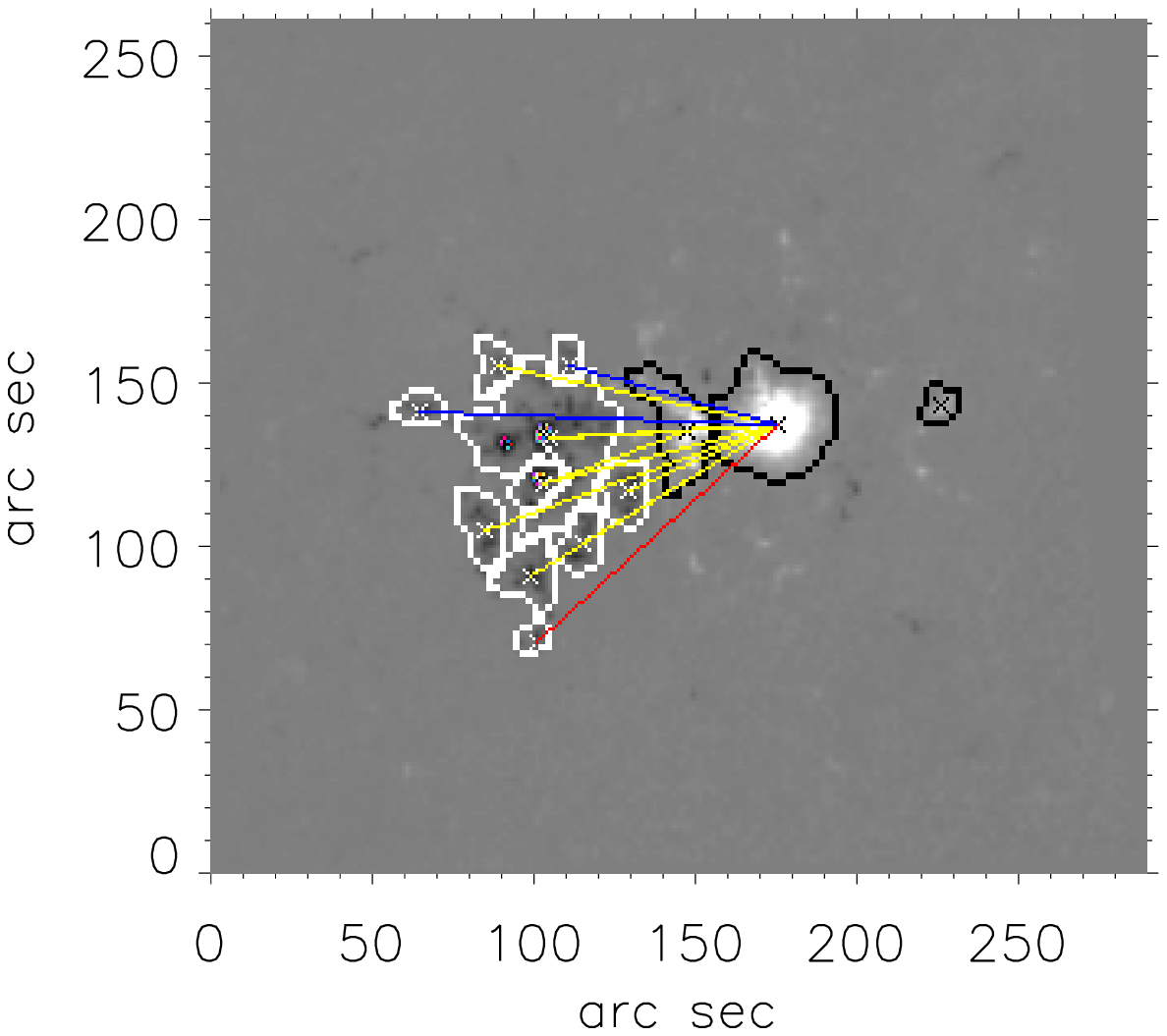}
\end{minipage}%
\begin{minipage}[t]{0.48\textwidth}
\vspace{0pt}
\centering
\includegraphics[width=\textwidth,height=0.8\textwidth]{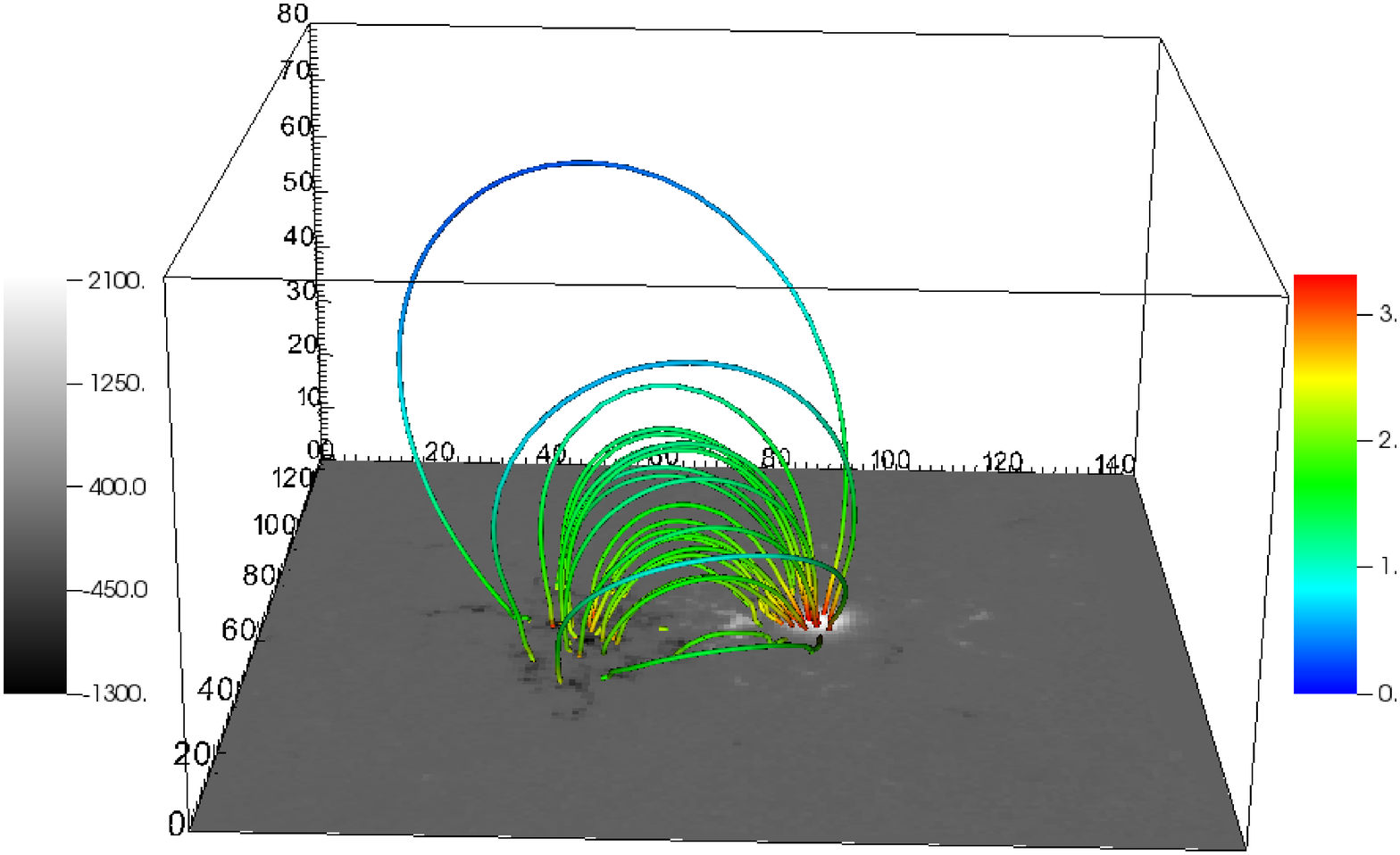}
\end{minipage}
\caption{{\bf Left panel:} Inferred magnetic connectivity of the non-eruptive NOAA AR 11072, observed on 2010 May 24 at 06:10 UT (see Section~\ref{S-obs}), showing the vertical magnetic field component in grayscale with the contours bounding the identified magnetic partitions. Line segments represent flux-tube connections, identified by the magnetic connectivity matrix, closing within the field-of-view and connecting the flux-weighted centroids of the respective pair of partitions. Red, blue, yellow, and cyan segments denote magnetic flux contents within the ranges $<5\times10^{19}$ Mx, [5$\times10^{19}$, 10$^{20}$] Mx, [10$^{20}$, 5$\times 10^{21}$] Mx, and $>5\times 10^{21}$ Mx, respectively. {\bf Right panel:} Three-dimensional extrapolated structure of the magnetic field, with magnetic field lines coloured according to the logarithm of the magnetic field strength. Spatial scales are in units of pixels.}\label{figdata3}
\end{center}
\end{figure}

For NOAA AR 11158 we use 30 vector magnetograms covering a 5-day period (2011 February 12-16) with a 4-hour cadence. This AR showed significant flaring and eruptive activity, releasing one X-class, 3 M-class and 25 C-class flares over the 5-day observing interval, and has hence been extensively studied in literature (see \opencite{tgl13}, and references therein). The photospheric area covered by these magnetograms is 300$\times$300 pixels, or 300$\times$300 arcsec on the image plane (pixel size of $\sim1$ arcsec). These magnetograms, in particular, were disambiguated and de-projected as described by \inlinecite{sun12}. We used the same single-value 1~$\sigma$-uncertainties as in the case of NOAA AR 11072.

\begin{figure}
\begin{center}
\begin{minipage}[t]{0.48\textwidth}
\vspace{0pt}
\centering
\includegraphics[width=\textwidth]{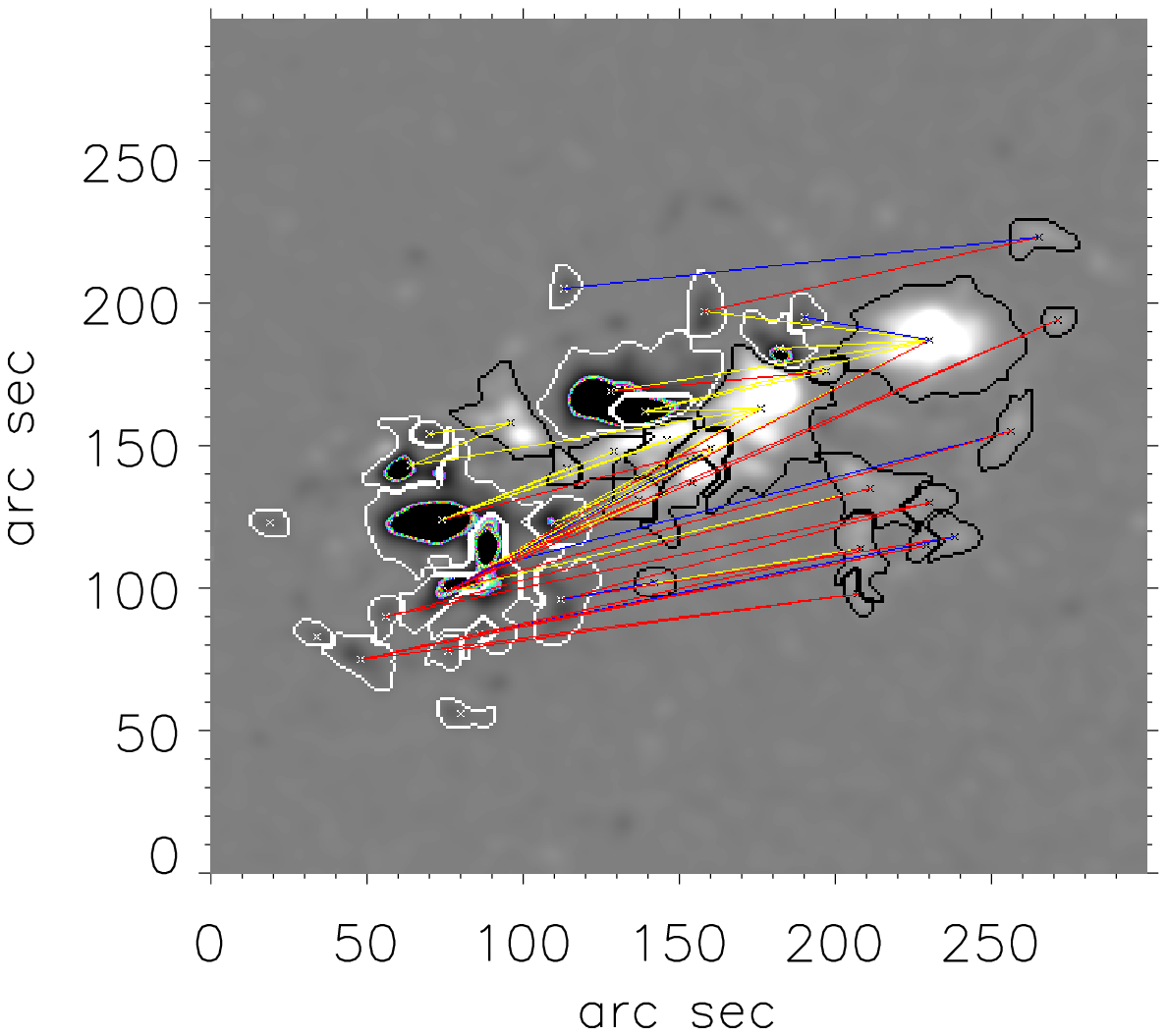}
\end{minipage}%
\begin{minipage}[t]{0.48\textwidth}
\vspace{0pt}
\centering
\includegraphics[width=\textwidth,height=0.8\textwidth]{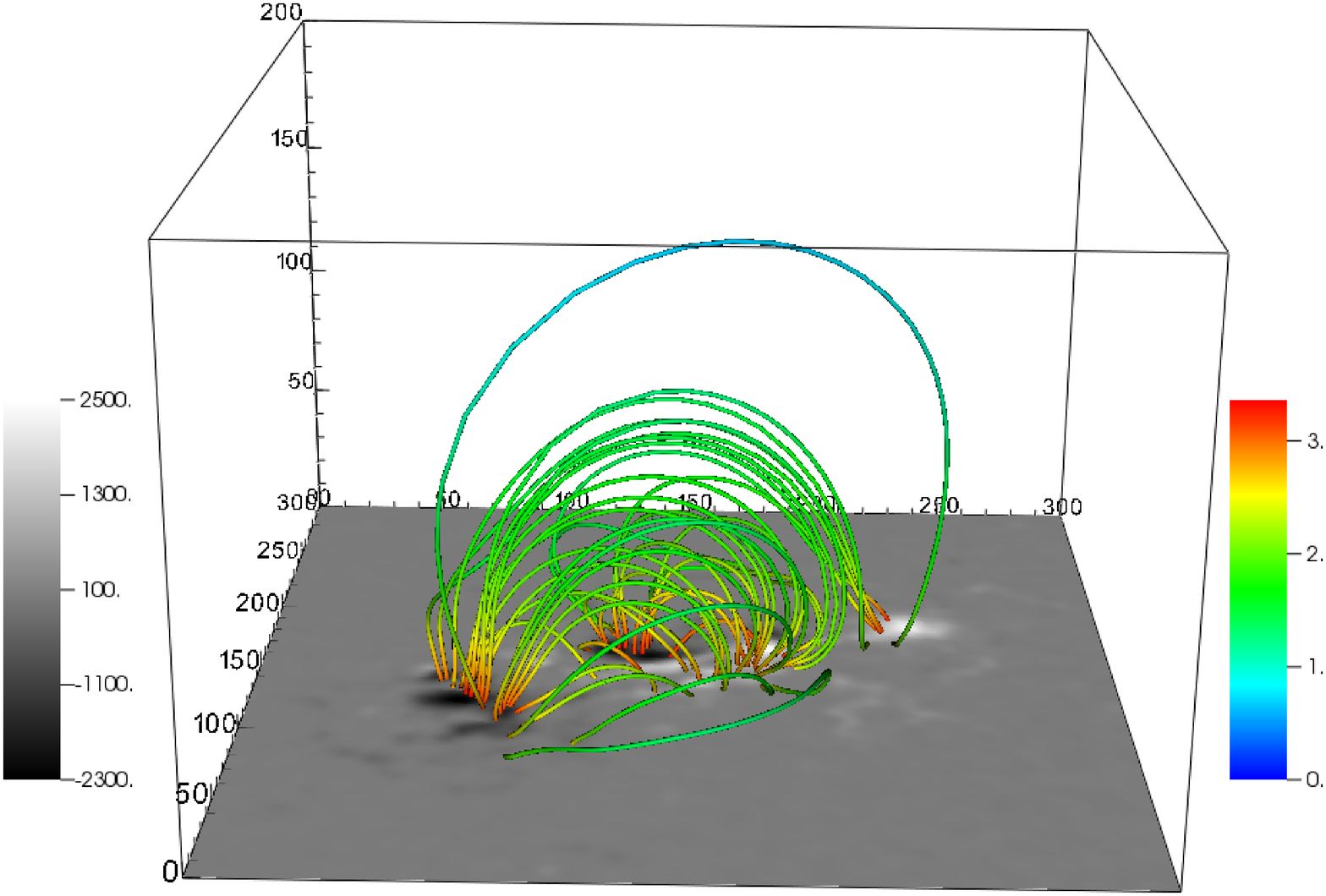}
\end{minipage}
\caption{Same as Figure~\ref{figdata3}, but for the eruptive NOAA AR 11158, observed on 2011 February 15 at 08:00 UT.}\label{figdata4}
\end{center}
\end{figure}

For our analysis we employ nonlinear force-free field extrapolations of the aforementioned time series of magnetograms of NOAA ARs 11072 and 11158, derived using the extrapolation code of \inlinecite{wieg04}. The three-dimensional magnetic field for NOAA AR 11072, that was derived without prior pre-pro\-cessing of the vector magnetograms, covers a volume with an average size of $220\times190\times220$~Mm (pixel size of $\sim$1460~km), with each spatial dimension ranging between 190~Mm and 280~Mm due to the heliographic projection used, while for NOAA AR 11158, where pre-processing was used, it covers a volume of $216\times216\times184$~Mm (pixel size of $\sim$720~km).

In Figures~\ref{figdata3}, \ref{figdata4} we show a snapshot of the three-dimensional structure of the magnetic field along with the NLFF-inferred photosperic magnetic connectivity, for the observed NOAA ARs 11072 and 11158 respectively.

\section{Results}
\label{S-results}

We use the NLFF and semi-analytical approaches described in Sections~\ref{S-equations} and \ref{S-simple-equations}, respectively, to derive the instantaneous budgets for the free magnetic energy and the relative magnetic helicity of the eruptive and non-eruptive synthetic MHD cases (Section~\ref{S-text}) and the respective observational cases (Section~\ref{S-obs}). For all cases we have also calculated the various flux budgets of interest (top plots of Figures~\ref{figa1}, \ref{figb1}, \ref{figd1}, \ref{figc1}). In particular, we show the total unsigned magnetic flux (black curves), the unsigned partitioned flux (i.e., total flux included in the identified magnetic partitions; magenta curves), and the unsigned connected flux (i.e., total flux included in the magnetic connectivity matrix; brown curves). Naturally, the flux budget of a given category is generally smaller than, and should be at most equal to, the flux of the previous category, with the total unsigned flux being the total flux included in the field of view and hence the upper limit of all flux budgets at any given time.

\subsection{Synthetic MHD cases}
\label{S-results-mhd}

\subsubsection{Non-eruptive case}
\label{S-results-nemhd}

\begin{figure}
\includegraphics[width=\textwidth]{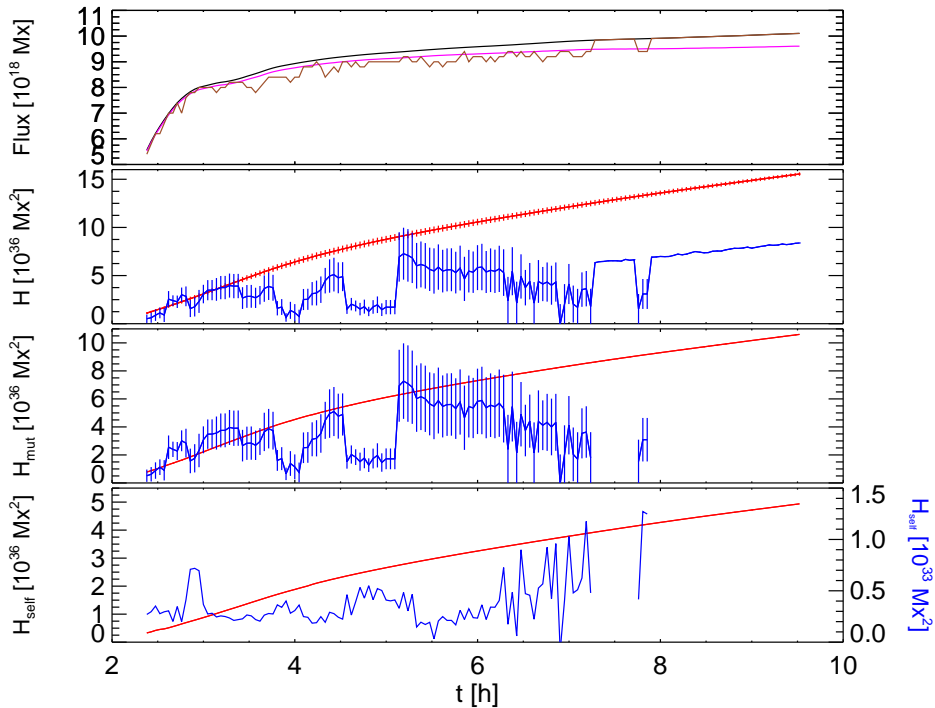}
\caption{Evolution of the total (black curve), partitioned (magenta curve) and connected (brown curve) unsigned flux (top panel), total relative helicity (second panel), mutual helicity (third panel) and self helicity (bottom panel) for the non-eruptive MHD case. Helicities obtained with the semi-analytical method are denoted with red color, while NLFF-derived helicities with blue. NLFF-derived self helicity has the units shown in the right axis, and is also plotted without errors since these are too large (for the chosen value range see text). Missing values in the bottom two panels correspond to LFF calculations, when the structure was deemed as simple as a single dipole, hence without mutual and self terms.}\label{figa1}
\end{figure}

\begin{figure}
\includegraphics[width=\textwidth]{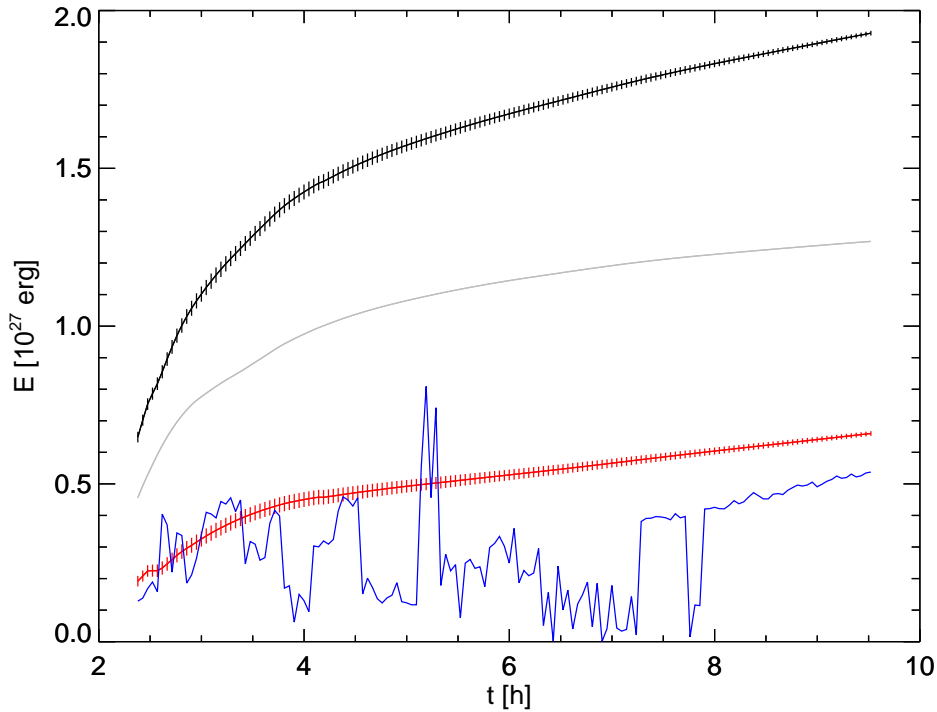}
\caption{Total (black curve), potential (gray curve) and free magnetic energy evolution for the non-eruptive MHD case. The semi-analytical and NLFF-derived free magnetic energies are respectively denoted with red and blue curves. Errors for the NLFF-derived free energy are large and are hence excluded for clarity (see text).}\label{figa2}
\end{figure}

For the synthetic non-eruptive case of Figure~\ref{figdata1}, we focused on times $t \gtrsim 2.4$~h to achieve more significant peak values of the magnetic field strength in the lower boundary ($B \gtrsim 30$~G). Figure~\ref{figa1} depicts the comparison results for the relative magnetic helicity, while the comparison for the respective magnetic energy budgets is shown in Figure~\ref{figa2}. For times $t\gtrsim 7.2$~h the synthetic AR has the form of a single dipolar structure, and there the LFF method is employed instead. This method uses practically the entire unsigned flux in the field of view as connected flux. Additionally, there are no self and mutual terms for the helicity in this case and the errors for the relative helicity are significantly smaller in the semi-analytical method than in the NLFF method.

In general, from Figures~\ref{figa1} and \ref{figa2} we notice that the volume-calculation method applied to the synthetic three-dimensional data shows a smooth increase of all budgets that is due to the smooth evolution of the synthetic structure, dominated by magnetic flux emergence (Figure~\ref{figa1}; upper panel). The results of the NLFF method show a generally increasing trend but they are much less smooth and more scattered. This is due to the fact that the NLFF method depends sensitively on the connected magnetic flux: the connected flux indeed shows a generally increasing trend, but with high-frequency variations superposed. This is the result of variations in the partitioning of the magnetic flux as, in this case, new emerging magnetic fields are weak and comparable to the magnetic field threshold values used for defining the magnetic partitions. This jiggling is amplified in the results for the free energy and the relative helicity. Uncertainties are hence significant for the NLFF method, particularly for this case that features a weak magnetic structure, with small free-energy and relative-helicity budgets compared to the eruptive case of Section~\ref{S-results-ermhd}. Errors are large in the case of NLFF-derived self helicity, as can be seen by its time average, $H_\mathrm{self}=(3.8\pm2.4)\times10^{32}\,\mathrm{Mx}^2$, and are thus supressed in Figure~\ref{figa1} for clarity.

In addition, Figures~\ref{figa1} and \ref{figa2} clearly show that, within uncertainties, the free-energy and relative-helicity budgets are smaller in the NLFF-calculation method than the respective budgets of the semi-analytical volume-calculation method. This result is expected and is additional evidence to the consistent construction of the NLFF method that infers a lower limit of the free magnetic energy and the corresponding relative magnetic helicity by ignoring inter-winding between individual flux tubes. Quantitative similarities and differences between the semi-analytical and NLFF relative-helicity and free-energy budgets are provided in Table~\ref{tab1}, where for each comparison we include the linear (Pearson) correlation coefficient, $r$, the rank-order (Spearman) correlation coefficient, $R$, and the ratio of average semi-analytical- to average NLFF-derived values, $f$, along with its 95\% confidence interval. 

Focusing on relative helicity budgets (Figure~\ref{figa1}), we notice (i) an agreement between the volume- and NLFF-calculation methods in terms of helicity sign (a right-handed magnetic structure inferred by both methods), and (ii) a significant-to-high correlation when the total relative helicity is concerned (correlation coefficients $\sim$ 0.72 - 0.76). This good correlation deteriorates in case one compares the mutual- or self-terms independently (0.35 - 0.38 for mutual; 0.29 - 0.34 for self terms, respectively). In addition, the self terms of the NLFF relative helicity are found to be at least three orders of magnitude smaller than those of the volume-calculated relative helicity. This striking result reflects the fact that self terms, in particular, are subject to the spatial resolution or, in our case, to the selected partitioning and subsequent number of flux tubes. The actual three-dimensional MHD model includes many more flux tubes than the abstract NLFF model. Generalizing, the gauge-invariant relative magnetic helicity shows an interplay between its mutual- and self-terms that is resolution-dependent. A consistent conclusion coming from this and previous studies, nonetheless, is that regardless of actual values, the mutual-helicity terms show larger to much larger amplitudes than the self-helicity terms (see, e.g., \opencite{regnier05}; \opencite{tgl13}).

Focusing on the free energy budgets (Figure~\ref{figa2}) we notice that both methods give values within the same range ($\lesssim 8 \times 10^{26}$~erg), but there is a weak correlation in terms of the temporal evolution of the free energy, resulting in generally low correlation coefficients (0.26 - 0.38). Besides the core assumptions of the NLFF method (zero Gauss linking number etc.), this weak correspondence can also be partly attributed to the small free-energy budgets in this particular example. For the NLFF method, these weak values give rise to relatively large uncertainties (the time average of NLFF-derived free energy is $E_\mathrm{c}=(3.1\pm1.6)\times10^{26}\,\mathrm{erg}$ and errors are omitted in Figure~\ref{figa2} for clarity), making the volume-calculated and NLFF free energies mostly similar within error bars.

\subsubsection{Eruptive case}
\label{S-results-ermhd}

\begin{figure}
\includegraphics[width=\textwidth]{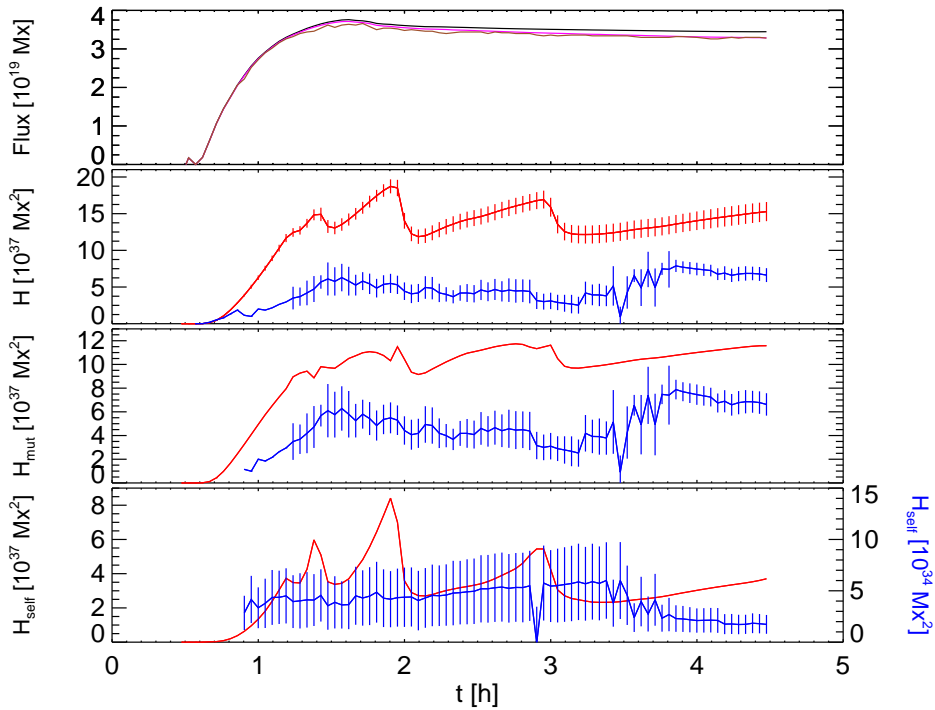}
\caption{Same as Figure~\ref{figa1}, but for the eruptive MHD case.}\label{figb1}
\end{figure}

\begin{figure}
\includegraphics[width=\textwidth]{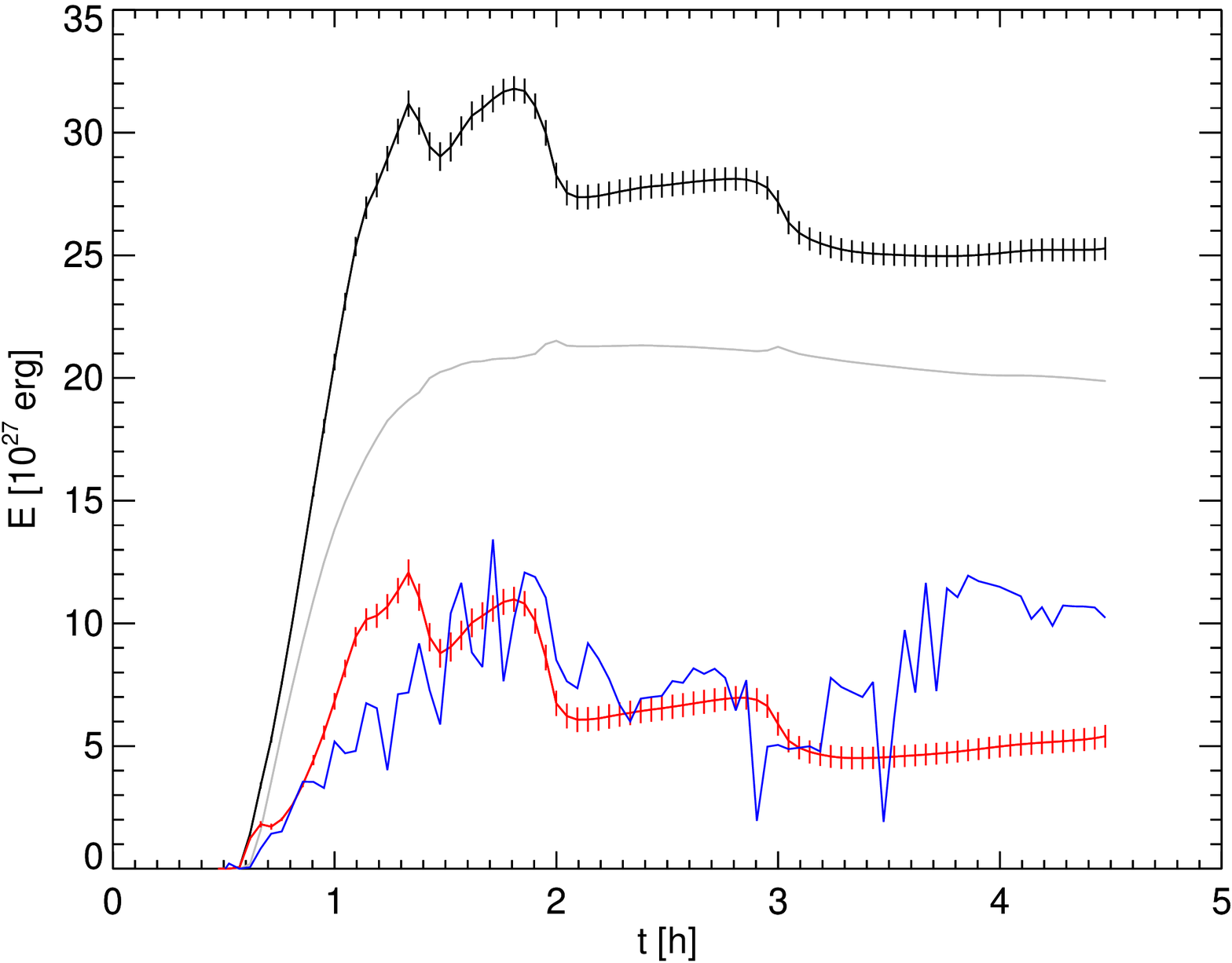}
\caption{Same as Figure~\ref{figa2}, but for the eruptive MHD case.}\label{figb2}
\end{figure}

For the synthetic eruptive case of Figure~\ref{figdata2}, we focused on times $t \gtrsim 0.5$~h to achieve significant field strengths ($B \gtrsim 30$~G). Figure~\ref{figb1} demonstrates the comparison results for the relative magnetic helicity while the respective comparison for the magnetic energy budgets is shown in Figure~\ref{figb2}.

The key difference between this and the non-eruptive case of Section~\ref{S-results-nemhd} is that at three different instances (t $\sim$ 1.4, 1.9, and 3~h) part of the magnetic structure is ejected outside the finite simulation volume. This appears clearly in the ``saw-tooth'' profile of the total relative helicity and the free-energy timeseries calculated by the semi-analytical volume-calculation method. Indeed, relative-helicity and free-energy budgets decrease as a result of the eruptions carrying part of the magnetic structure beyond the simulation volume. 

In this case the total unsigned, partitioned, and connected fluxes seem to show a better correspondence (Figure~\ref{figb1}; upper panel), at higher values than in the non-eruptive case. In addition, it is now clear that ignoring the inter-winding between flux tubes results in clearly smaller relative-helicity and free-energy budgets in the NLFF method - see also Table~\ref{tab1}.

Focusing on relative-helicity budgets (Figure~\ref{figb1}) the two methods also agree on the sign of the helicity (a right-handed magnetic structure inferred by both methods). In addition, the correlation coefficients between the two total relative helicities are significant-to-high (0.60 - 0.74). As with the non-eruptive case, discrepancies mainly correspond to the temporal profiles: the NLFF method generally fails to show a distinct signal at the time of the eruptions. Since the method depends sensitively on the connectivity matrix in the lower boundary, lack of a clear-cut response in the flux profile at this boundary due to eruptions will result in a lack of signal in the results of the method. From the flux profile (upper panel) we notice that this is, indeed, the case. Any likely changes in the relative helicity in the course of the eruptions are well within uncertainties, and hence not significant. 

As with the non-eruptive case, correlations weaken when the self- and mutual-helicity terms are considered (bottom two panels of Figure~\ref{figb1}). Given that self terms between the two methods practically show no correspondence, the mutual terms correlate better than in the non-eruptive case. In case of the self terms, the NLFF method again provides budgets that are $\sim 10^3$ times smaller than those of the volume-calculation method. 

Focusing on the free-energy budgets (Figure~\ref{figb2}), we notice that the values provided by the two methods are quite close, although temporal profiles again show discrepancies. Errors in NLFF-derived free energy are again large, the time average of free energy is $E_\mathrm{c}=(7.5\pm3.1)\times10^{27}\,\mathrm{erg}$, and are hence omitted from Figure~\ref{figb2}. This results in generally weak correlations (coefficients $\sim$ 0.28 - 0.43). One might argue that here, as well, the NLFF free-energy budget seems to decrease in the course of the eruptions in a manner similar to that of the volume-calculation budget. Regardless, these apparently eruption-related changes are well within applicable uncertainties, and are hence not significant. 

On average, for the above two MHD models, lack of knowledge of the three-dimensional magnetic structure in the NLFF-calculation method results in relative helicities that are $\sim(2.5\pm0.1)$ times smaller than the ``ground-truth'' semi-analytical, volume-calculated relative helicities. For the free energy, the NLFF method underestimates the respective semi-analytical values by a factor of $\sim(1.25\pm0.08)$. We also consider this an acceptable result given the simplicity and general applicability of the NLFF method. 

The results of the above comparisons are summarized in Table~\ref{tab1}.


\begin{table}
\caption{Comparison between results for the synthetic MHD cases} 
\resizebox{\textwidth}{!}{

\centering
\begin{tabular}{@{}lccccccccccccccc@{}}
\toprule
 & \multicolumn{3}{c}{$H$} & \phantom{ab} & \multicolumn{3}{c}{$H_\mathrm{mut}$} & \phantom{ab} & \multicolumn{3}{c}{$H_\mathrm{self}$} & \phantom{ab} & \multicolumn{3}{c}{$E_\mathrm{c}$} \\ 
\cmidrule{2-4} \cmidrule{6-8} \cmidrule{10-12} \cmidrule{14-16}
 & $r$ & $R$ \tabnote{$r$: linear correlation coefficient, $R$: rank-order correlation coefficient} & $f$ \tabnote{$f$: ratio of average semi-analytical to average NLFF-derived value} && $r$ & $R$ & $f$ && $r$ & $R$ & $f$ && $r$ & $R$ & $f$ \\ 
\midrule
non-eruptive & 0.72 & 0.76 & $2.11^{+0.13}_{-0.11}$ && 0.38 & 0.35 & $1.96^{+0.22}_{-0.19}$ && 0.34 & 0.29 & $(8.0^{+1.1}_{-0.9})$ && 0.26 & 0.38 & $1.66^{+0.15}_{-0.13}$ \\
 & & & &&& & & & & &$\times10^3$&& & & \\ 
eruptive & 0.74 & 0.6 & $2.83^{+0.21}_{-0.19}$ && 0.6 & 0.48 & $1.91^{+0.20}_{-0.18}$ && 0.062 & -0.007 & $(7.8^{+1.1}_{-1.0})$ && 0.43 & 0.28 & $0.85^{+0.09}_{-0.08}$ \\
 & & &&&& & & & & &$\times10^2$&& & & \\ 
\bottomrule
\end{tabular}
}
\label{tab1}
\end{table}

\subsection{Observational cases}
\label{S-results-obs}

As explained in Section~\ref{S-obs}, comparison between the semi-analytical volume-calculation and NLFF method results in observed solar active-region cases has to rely on nonlinear force-free field-extrapolations that are necessary for the volume-calculation method. As extrapolations are also known to include uncertainties and model-dependencies \cite{schrijver06,metc08}, possible discrepancies will need to be viewed under the prism of errors and uncertainties stemming from both methods.

\subsubsection{Non-eruptive NOAA AR 11072}

\begin{figure}
\includegraphics[width=\textwidth]{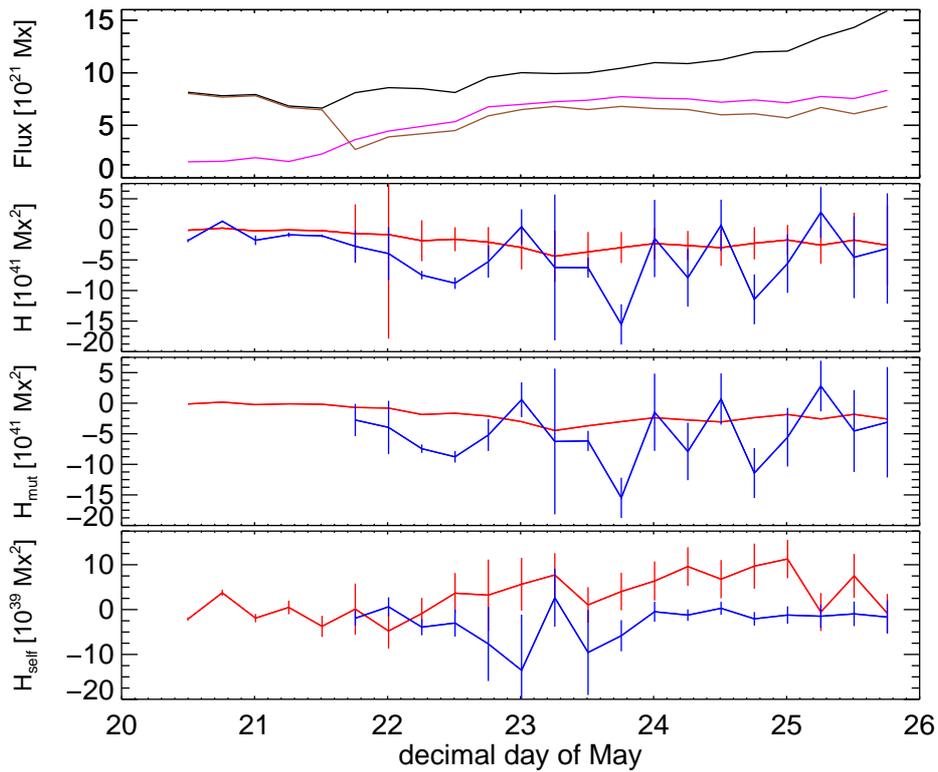}
\caption{Same as Figure~\ref{figa1}, but for the observed non-eruptive NOAA AR 11072 in 2010 May.
}\label{figd1}
\end{figure}

For the non-eruptive observed NOAA AR 11072 of Figure~\ref{figdata3}, Figure~\ref{figd1} (upper panel) shows the total unsigned, partitioned, and connected fluxes over a $\sim$5-day period in 2010 May. In this case, patches of quiet-Sun or otherwise weak (thus not included in the partitioning) flux also appear in the field-of-view, forcing the partitioned flux of the NLFF method not to exceed $\sim$65\% of the total unsigned flux on average. For the first five snapshots (from start to May 21 at 12:10 UT) the AR shows a single dipolar structure and the LFF method is used instead.

The total relative helicity in the AR is rather small (i.e., peaking at $\sim -1.5 \times 10^{42}\,\mathrm{Mx}^2$ for the NLFF method and at $\sim -5 \times 10^{41}\,\mathrm{Mx}^2$ for the volume-calculation method) but predominantly left-handed, as inferred by both methods. The volume-calculation method results in consistently lower and relatively smoother amplitudes than the NLFF method. The correlation coefficients are therefore low (0.31 - 0.35) and they become even lower when the mutual and self terms of helicity are compared independently. Remarkably, though, correlations in this case seem better for the self-terms than for the mutual terms, contrary to the results of synthetic MHD cases. 

Another worth-mentioning finding is that the volume-calculated self helicity seems predominantly right-handed, that is, of opposite sense to the total relative helicity, at least from 2010 May 22 onwards. The NLFF method gives consistently left-handed total helicity and like-sense self and mutual terms. If the volume-calculated result is to be trusted, then this warrants additional investigation as one intuitively expects that the signs of mutual- and self-helicity should agree overall, particularly in view of the non-ideal, magnetic-reconnection-powered interplay between the two, enabled by the conversion of mutual-to-self helicity \cite{tgl13}. Point taken, we cannot rule out that this feature might be due to uncertainties in the extrapolation. 

\begin{figure}
\includegraphics[width=\textwidth]{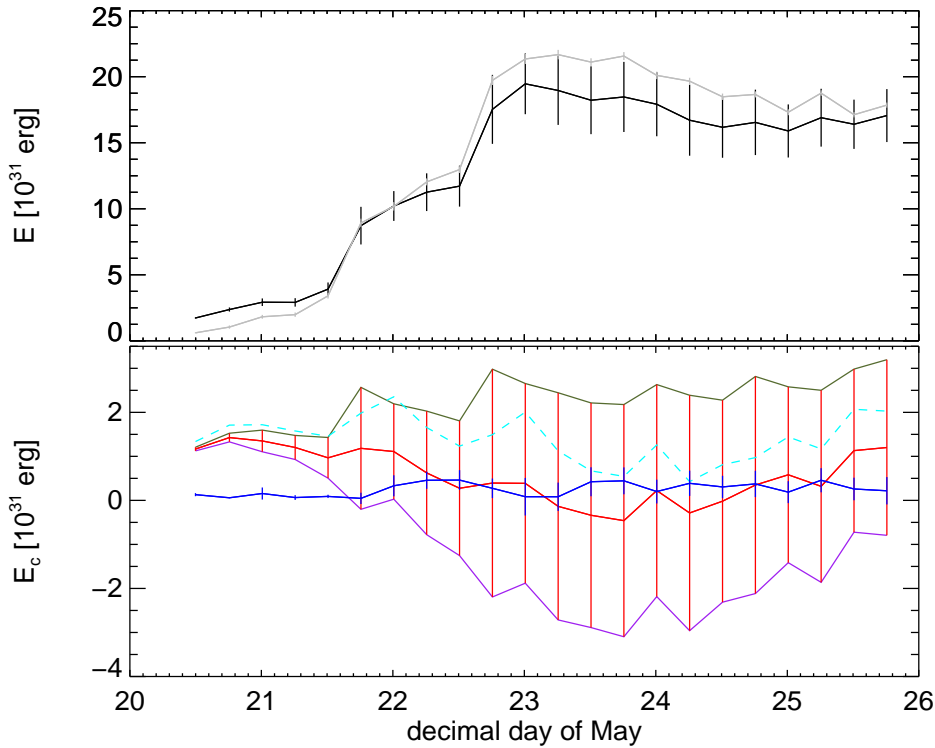}
\caption{Same as Figure~\ref{figa2}, but for the observed non-eruptive NOAA AR 11072 in 2010 May. The purple and pale green limiting curves in the bottom panel correspond to the (theoretically equivalent) expressions of Equations~(\ref{ecdef}) and (\ref{ecdef2}), respectively, for the magnetic free energy, while the cyan curve is the free energy obtained with the Schmidt potential.}\label{figd2}
\end{figure}

Additional questions on the validity of the field extrapolation are borne out of the comparison between the free magnetic energy budgets (Figure~\ref{figd2}). There, it is clear that the free energy $E_\mathrm{c}$ from Equation~(\ref{ecdef}) becomes negative after 2010 May 22, giving rise to large uncertainties when averaged with the always positive, theoretically equivalent, free energy $E'_\mathrm{c}$ from Equation~(\ref{ecdef2}). \inlinecite{val13} warned that negative free energies may arise as results of extrapolations where the divergence-free condition has not been satisfactorily achieved. We agree that this is clearly an issue with the extrapolation, resulting to unphysical free energies. This is also supported by the value of the volume-averaged, absolute fractional flux increase \cite{wheat00}, which varies in the range $<|f_i|>=(1.34\pm0.24)\times10^{-3}$ during the evolution of the AR. The corresponding current weighted angle between $\mathbf{J}$ and $\mathbf{B}$ is $\theta_J=26^\circ.1\pm1^\circ.6$, indicating a rather poorly constructed force-free field. We should note however that negative free energies are avoided (see bottom panel of Figure~\ref{figd2}) when we consider the potential magnetic field in the semi-infinite volume above the lower boundary as in \inlinecite{schm64}. Then one finds the absolutely minimum potential energy for the given lower-boundary condition and the derived free energy is positive, however the two implemented volumes are not equivalent to each other. 

The NLFF free energy shows less scatter in this case, is well within uncertainties of the volume-calculated free energy, and peaks at $10^{31}\,\mathrm{erg}$ - this is a particularly low value corroborating the non-eruptive, but also flare-quiet, nature of NOAA AR 11072. These problems in the volume-calculated free energy result in an anti-correlation between the two free energy budgets, with correlation coefficients in the range (-0.57, -0.58). It becomes evident that extrapolation results in this case should be handled with extreme caution. 

The use of preprocessing \cite{wieg06} deteriorates the situation in this case, resulting in even lower free-energy values when the \inlinecite{schm64} potential field is used. We speculate that this could be attributed to the non-eruptive character of the specific AR, with such small free energy budgets, and perhaps, to failing to satisfy the criteria for a successful extrapolation \cite{derosa09} regarding the size of the volume and the resolution.


\subsubsection{Eruptive NOAA AR 11158}

\begin{figure}
\includegraphics[width=\textwidth]{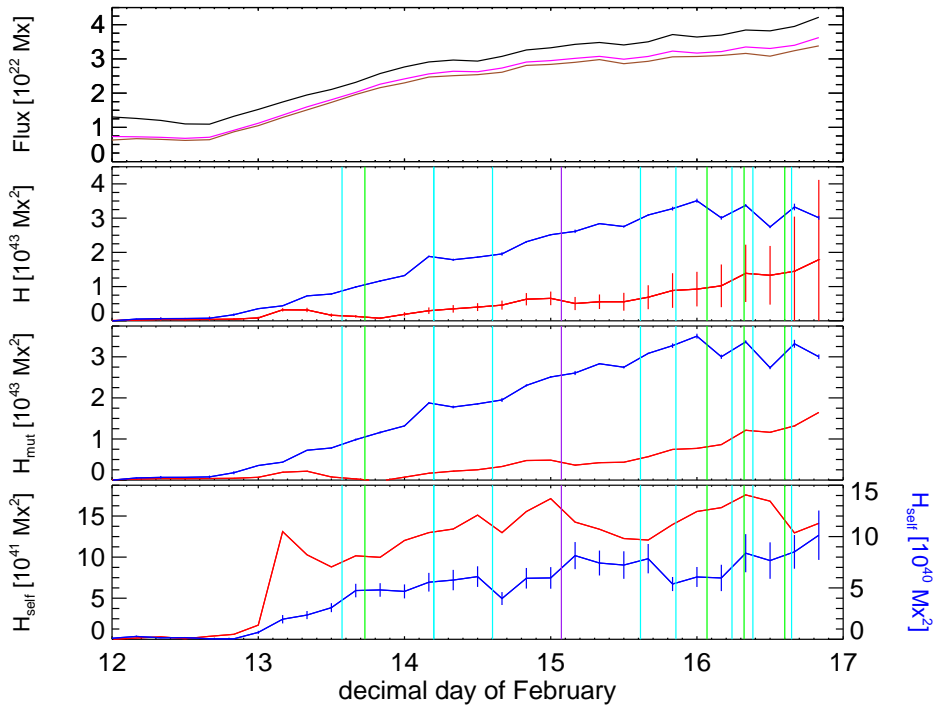}
\caption{Same as Figure~\ref{figa1}, but for the observed eruptive NOAA AR 11158 in 2011 February. Vertical lines mark the peak times of all flares greater than C4.0 observed during this five-day observing interval. The purple line corresponds to the main X2.2 flare, while green and cyan lines correspond to M- and C-class flares, respectively.
}\label{figc1}
\end{figure}

For the eruptive NOAA AR 11158 of Figure~\ref{figdata4}, Figure~\ref{figc1} (upper panel) shows the total unsigned, partitioned, and connected fluxes over a five-day period in 2011 February. Due to quiet-Sun or otherwise weak flux patches not included in the partitioning, the partitioned flux peaks, on average, at $\sim$80\% of the total unsigned flux. Practically all partitioned flux is also connected flux, participating in the magnetic-connectivity matrix and hence in the NLFF method calculation. 

The total relative helicity in the AR is found to be predominantly right-handed by both methods, as also inferred by scores of previous studies. The same is the case for the mutual- and the self-helicity terms, as well. In addition, all helicity terms show an increasing tendency, albeit with different increase rates for the two methods, owning to the increasing flux budgets stemming from a continuous flux emergence in the AR. Two distinct findings from Figure~\ref{figc1} are that (i) the correlation coefficients between the two methods are quite high, in the range (0.84 - 0.94) for the total relative helicity, (0.80 - 0.90) for its mutual term, and (0.77 - 0.86) for its self term, and (ii) the NLFF method consistently gives higher budgets than the volume-calculation method. Differences well exceed applicable uncertainties. Both findings seem to run counter to findings for the synthetic MHD cases. In addition, we emphasize that the NLFF method is designed to provide a minimum free energy and the corresponding relative helicity. This result, therefore, seems to defy basic principles of the NLFF method construction.

\begin{figure}
\includegraphics[width=\textwidth]{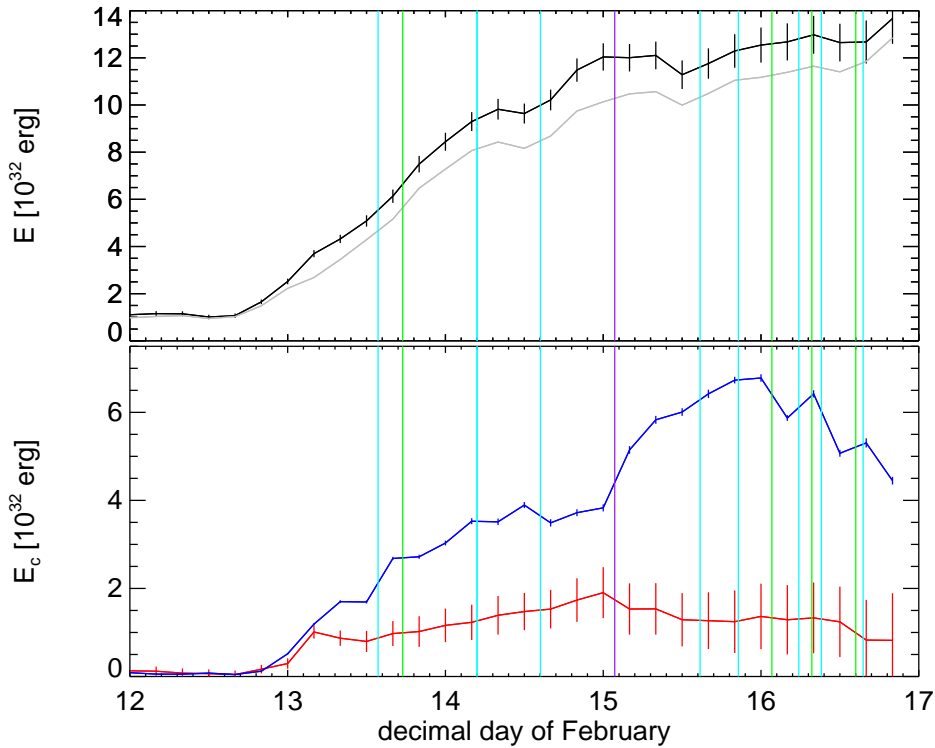}
\caption{Same as Figure~\ref{figa2}, but for the observed eruptive NOAA AR 11158 in 2011 February. Vertical lines are as in Figure~\ref{figc1}.}\label{figc2}
\end{figure}

The situation is similar when the free-energy budgets for the two methods are compared (Figure~\ref{figc2}). Besides overall significant correlation coefficients (0.72 - 0.78), the NLFF free energy peaks at $\sim 7 \times 10^{32}$~erg, while the volume-calculated free energy peaks at $\sim 1.5 \times 10^{32}\,\mathrm{erg}$ and has a time-profile similar to the one reported in \inlinecite{sun12}. Here the extrapolation results behave well, with $E_\mathrm{c}$ and $E'_\mathrm{c}$ being both positive and close to its other, as can be inferred by the magnitude of the errors of the volume-calculated free energy, and additionally, by the average value of the fractional flux increase $<|f_i|>=(7.2\pm0.9)\times10^{-4}$, which is half of that of AR 11072. Perhaps this is due to the substantial free-energy budget of the AR, more than an order of magnitude larger than in the non-eruptive NOAA AR 11072. Per the assessment of \inlinecite{val13}, therefore, the extrapolated field is reasonably divergence-free. Nonetheless, we argue at this point that the higher NLFF free energies compared to their volume-calculated counterparts, as well as the higher respective relative-helicity budgets, are again due to the performance of the extrapolation. This claim will be discussed in Section~\ref{S-discuss} at some detail. 

On average, for the above two observed ARs, lack of knowledge of the three-dimensional magnetic structure in the NLFF calculation method results in relative helicities that are $\sim(2.7\pm0.9)$ times larger than the semi-analytical, volume-calculated relative helicities. For the free energy, the NLFF method overestimates the respective volume-calculated values in the case of NOAA AR 11158 by a factor of $\sim3.3$ while in the case of NOAA AR 11072 it underestimates it by a similar factor, $\sim2.3$, and thus no meaningful average can be obtained. Contrary to the synthetic MHD models of Section~\ref{S-results-mhd}, however, model-dependent extrapolation results and subsequent findings cannot be considered as ``ground truth''. The purpose of this exercise is to practically assess how close the NLFF-calculation results are to results relying on a generally accepted, widely used nonlinear force-free field extrapolation method. Of course, our conclusions pertain to our application of the chosen field extrapolation method. Applications with significantly different extrapolated fields might lead us to significantly different findings.

Table~\ref{tab2} summarizes the results of the above comparisons for the observed AR cases.

\begin{table}
\caption{Comparison between results for the observational cases} 
\centering
\resizebox{\textwidth}{!}{
\begin{tabular}{@{}lccccccccccccccc@{}}
\toprule
 & \multicolumn{3}{c}{$H$} & \phantom{ab} & \multicolumn{3}{c}{$H_\mathrm{mut}$} & \phantom{ab} & \multicolumn{3}{c}{$H_\mathrm{self}$} & \phantom{ab} & \multicolumn{3}{c}{$E_\mathrm{c}$} \\ 
\cmidrule{2-4} \cmidrule{6-8} \cmidrule{10-12} \cmidrule{14-16}
 & $r$ & $R$ & $f$ \tabnote{Symbols are as in Table~1} && $r$ & $R$ & $f$ && $r$ & $R$ & $f$ && $r$ & $R$ & $f$ \\ 
\midrule
AR 11072 & 0.35 & 0.31 & $0.45^{+0.34}_{-0.15}$ && 0.051 & -0.022 & $0.37^{+0.33}_{-0.15}$ && 0.11 & 0.26 & *\tabnote{No meaningful value can be obtained in this case} && -0.57 & -0.58 & $2.3^{+1.9}_{-1.2}$ \\
AR 11158 & 0.84 & 0.94 & $0.30\pm0.06$ && 0.8 & 0.9 & $0.24\pm0.06$ && 0.86 & 0.77 & $23^{+3}_{-2}$ && 0.78 & 0.72 & $0.30^{+0.06}_{-0.04}$ \\
\bottomrule
\end{tabular}
}
\label{tab2}
\end{table}

\section{Discussion and Conclusions}
\label{S-discuss}

This work aims to address a largely outstanding question pertaining to the calculation of magnetic energy and helicity budgets in observed solar active regions: how close to, or far from, ``ground truth'' can estimates that lack vital information on the detailed three-dimensional coronal magnetic structure of active regions be? This question is both important and timely, as high-quality, but inherently photospheric, vector magnetogram data from SDO/HMI and other state-of-the-art instruments are accumulated at an accelerated pace. 

To this purpose, we have jointly used three-dimensional MHD models and observed active-region magnetograms. For the synthetic, MHD active-region cases, the ``ground truth'' coronal magnetic field is fully known. For the photospheric active-region magnetograms we relied on a widely used, established nonlinear force-free extrapolation method hoping to achieve some insight of the coronal ``ground truth''. Three-dimensional information available, we utilized the semi-analytical, volume-based magnetic free-energy and relative-helicity calculation method of Section~\ref{S-simple-equations}. In addition, we used the numerical, boundary-based NLFF calculation method of Section~\ref{S-equations} to attempt the same energy and helicity calculations assuming lack of knowledge of the three-dimensional magnetic field. 

The main finding, regardless of MHD data or observed active-region magnetograms, is that volume- and boundary-calculated free magnetic energies and relative magnetic helicities differ by a factor of $\sim$3, at most (Tables~\ref{tab1}, \ref{tab2}). This pertains to total and mutual terms of the free energy and helicity, with the much smaller self-terms differing widely, sometimes by orders of magnitude (Table~\ref{tab1}). The large discrepancy for self-terms is generally expected and reflects the large differences in the numbers of flux tubes involved in the volume- and boundary-calculations in view of the dependence of self-helicity in spatial resolution. When total or mutual free-energy and relative-helicity budgets are concerned, nonetheless, our findings present a sound improvement over published comparisons between, say, the volume calculation and time-integrated energy and helicity injection rates that differ by an order of magnitude or more (see \inlinecite{tgl13} and references therein). Therefore, our numerical NLFF method can be used to assess at least the amplitude of free energy and relative helicity for practical, solar active-region applications, without the need to extrapolate for the unknown coronal magnetic field or solve for the photospheric flow velocity from magnetogram timeseries. Both of these tasks are computationally expensive and time-consuming, besides being susceptible to large uncertainties and model dependencies, as discussed in Section~\ref{S-Introduction}.

Despite relative amplitude similarities, however, our results show discrepancies in the temporal profiles of volume- and NLFF-derived free energies and helicities. These discrepancies increase as free energy and relative-helicity values become smaller, i.e. for the non-eruptive MHD and observed NOAA AR 11072 cases. On the contrary, for the eruptive NOAA AR 11158 we find a high correlation in terms of relative helicity (correlation coefficients $\sim$0.9) and a lower, but still significant, correlation in terms of free energy (correlation coefficients $\sim$0.75). Discrepancies may be due to (i) large uncertainties in the NLFF calculation or (ii)  the marginal, if any, lower-boundary response to eruptions, for example in the eruptive MHD case of this study. In high-cadence free energy and helicity timeseries of NOAA AR 11158, \inlinecite{tgl13} were able to identify changes that they interpreted as eruption-related. Discrepancies are obviously also due to the fact that NLFF-derived values are lower-limit values, with a generally nonlinear departure from ``ground-truth'' values, due to the unknown flux-tube braiding and linkage in the corona. At least for eruptive active-region cases with sizable free-energy and helicity budgets, however, our NLFF method is found to perform quite well. This is because the simulated-annealing magnetic connectivity approach of the NLFF method is precisely constructed to favor compact, eruptive active regions.  

As already explained, while the three-dimensional magnetic field information from MHD models can be considered ``ground truth'', this is not the case with the respective information provided by the model-dependent nonlinear force-free field extrapolations. In fact, the validation exercise of this study furnishes {\it two} independent quality tests for the results of any nonlinear force-free extrapolation method that optimizes and does not strictly enforce the validity of the divergence-free and the zero Lorentz-force conditions. For instance, the optimization method of \citeauthor{wieg04} (\citeyear{wieg04} and subsequent works) pursues the combined minimization of $|\nabla \cdot \mathbf{B}|$ and $|\mathbf{J} \times \mathbf{B}|$. In case $|\nabla \cdot \mathbf{B}|$ is not exactly zero, the two free-energy expressions of Equations~(\ref{ecdef}) and (\ref{ecdef2}) show discrepancies, as \inlinecite{val13} also warned. Evidence of this shortcoming in the extrapolation is evident for the case of NOAA AR 11072 (Figure~\ref{figd2}). In addition, the fact that our NLFF method infers a lower limit for the free magnetic energy is in stark contrast with the results for NOAA AR 11158, where it gives us $\sim3$ times larger free energy compared to the volume-calculated free energy{\footnote{Notice however that, in this case, the reasonable difference between $E_c$ and $E'_c$ indicates that the extrapolation has nonetheless achieved a near-zero $|\nabla \cdot \mathbf{B}|$.}} (Figure~\ref{figc2}; bottom plot). The reason for this apparent paradox could be attributed to the minimization of $|\mathbf{J} \times \mathbf{B}|$: this can be achieved either by constructing $\mathbf{B}$ such that $\nabla \times \mathbf{B} \sim \mathbf{J}$ is mostly parallel to $\mathbf{B}$, albeit not zero, or by pursuing $|\nabla \times \mathbf{B}| \sim 0$ , regardless of $\mathbf{J}$-orientation. Both practices may be formally valid, but the second will tend to give a near current-free, rather than a force-free, solution. Since we do not have any control on how $|\mathbf{J} \times \mathbf{B}|$ is minimized, we assume that the extrapolation in NOAA AR 11158 has acted to produce a nearly current-free solution, hence the result of Figure~\ref{figc2}. Partial support for this claim stems from the  current weighted angle between $\mathbf{J}$ and $\mathbf{B}$ that is in the range $\theta_J=17^\circ.3\pm0^\circ.9$, but more importantly, from the broad distribution of the individual angles between $\mathbf{J}$ and $\mathbf{B}$ with respect to the magnitude of the current. Similar tests may be performed to assess the quality of any given extrapolation results.

It is worth-mentioning that semi-analytical and NLFF-derived free magnetic energy and relative magnetic helicity budgets, for \textit{both} synthetic and observed active regions, do follow the monotonic scaling first reported by \inlinecite{tgr12} and independently confirmed for NOAA AR 11158 \cite{tgl13} and quiet-Sun regions \cite{tz14}. However, as this finding is out of the scope of the present analysis concerning validation of the NLFF method, this result and its physical implications will be further discussed in a forthcoming paper. 

Concluding, we argue that the NLFF method of \inlinecite{geo12a}, introduced to perform a practical, self-consistent calculation of magnetic free energy and relative magnetic helicity in (preferably eruptive) solar active regions without requiring photospheric flows or three-dimensional coronal fields, has been validated in a sufficient and generally successful manner in the framework of this study. In addition, we now possess a state-of-the-art method to apply conventional, semi-analytical volume-based calculations for magnetic free energy and helicity, should complete three-dimensional magnetic field data exist. Future juxtaposition and coupling between these two complementary tools, the former emphasizing accuracy and the latter devoted to computational efficiency, is envisioned to lead to further advances, understanding, and insight into the eruptive potential of solar active regions and the robust quantification of this potential.  

\appendix

\section{Error estimation for the semi-analytic method}
  \label{S-simple-equations-errors}

From the vector potentials $\mathbf{A}$, $\mathbf{A}_\mathrm{p}$ we define the reconstructed fields $\mathbf{B}^*=\nabla\times\mathbf{A}$, $\mathbf{B}_\mathrm{p}^*=\nabla\times\mathbf{A}_\mathrm{p}$. We then assign an uncertainty in each field simply as $\delta \mathbf{B}=\mathbf{B}-\mathbf{B}^*$ and $\delta \mathbf{B}_\mathrm{p}=\mathbf{B}_\mathrm{p}-\mathbf{B}_\mathrm{p}^*$. Of course, uncertainties and errors in our calculations have various origins, such as the solution of Laplace's equation or the numerical integrations for obtaining the vector potentials. However, since we cannot quantify these errors trivially, we simply treat them as zero and define a single error from the difference between the given and reconstructed fields, which can serve as a lower limit in the actual error.

The error in the total relative helicity can be written as
\begin{equation}
\delta H=\sqrt{(\delta H_{\rm mut})^2+(\delta H_{\rm self})^2},
\label{errh1}
\end{equation}
assuming that the two errors are independent. Since all quantities of interest are volume integrals of their respective density, that is, of the form $X=\int {\rm dV}\,\mathcal{X}$, the error in them will be
\begin{equation}
\delta X=\lambda^3 \sqrt{\sum_{i}(\delta \mathcal{X}_i)^2},
\label{errh2}
\end{equation}
where $\lambda^3$ represents the volume element and summation is over all points in the volume. Regarding the mutual helicity density error, standard error propagation leads to
\begin{equation}
\delta \mathcal{H}_{\mathrm{mut},i}=\left(4A_{px}^2(\delta B_x^2+\delta B_{px}^2)+4A_{py}^2(\delta B_y^2+\delta B_{py}^2)\right)^{1/2},
\label{errh3}
\end{equation}
where all quantities in the right hand side are to be taken at point $i$. In a similar fashion, we can find the error in self helicity density as
\begin{equation}
\delta \mathcal{H}_{\mathrm{self},i}=\left((A_x-A_{px})^2(\delta B_x^2+\delta B_{px}^2)+(A_y-A_{py})^2(\delta B_y^2+\delta B_{py}^2)\right)^{1/2}.
\label{errh4}
\end{equation}

Likewise, the error in free energy can be obtained from
\begin{equation}
\delta E_\mathrm{c}=\sqrt{(\delta E_{\rm t})^2+(\delta E_{\rm p})^2},
\label{errec1}
\end{equation}
where the error in total energy density is
\begin{equation}
\delta \mathcal{E}_{\mathrm{t},i}=\frac{1}{4\pi}\left((B_x\delta B_x)^2+(B_y\delta B_y)^2+(B_z\delta B_z)^2\right)^{1/2}.
\label{errec2}
\end{equation}
and a similar formula holds for the error in potential energy density, $\delta \mathcal{E}_{\mathrm{p},i}$, if we replace $\mathbf{B}$, $\delta \mathbf{B}$ by $\mathbf{B}_\mathrm{p}$, $\delta \mathbf{B}_\mathrm{p}$, respectively.

Another stringent error estimation stems from the difference between the two theoretically equivalent free-energy formulas of Equations~(\ref{ecdef}), (\ref{ecdef2}), i.e. the imperfect solenoidal property of $\mathbf{B}$. In this respect, an alternative error in free magnetic energy can be defined by $\delta E'_\mathrm{c}=\frac{1}{2}\left|E_\mathrm{c}-E'_\mathrm{c}\right|$ and then the final error will be $\max (\delta E_\mathrm{c},\delta E_\mathrm{c}')$. In the cases where $\delta E_\mathrm{c}'>\delta E_\mathrm{c}$, and in order to be consistent, the error in total energy will be recalculated from Equation~(\ref{errec1}) with $E_\mathrm{c}$, $E_\mathrm{t}$ interchanged.

Taking into account at this point the LFF energy-helicity study of \inlinecite{gbont07} and assuming their simplified relation between the free magnetic energy and the relative magnetic helicity, the error $\delta E'_\mathrm{c}$ in free energy gives rise to an error $\delta H'$ in the relative helicity of the form
\begin{equation}
\delta H'=H\frac{\delta E'_\mathrm{c}}{E_\mathrm{c}}.
\end{equation}
The final selection of the error for relative magnetic helicity will thus be taken as $\max (\delta H,\delta H')$, while the mutual- and self-helicity errors will be calculated by Equations~(\ref{errh2}), (\ref{errh3}) and (\ref{errh4}).

\section{Performance of the semi-analytic method in simulated data with known energy/helicity budgets}
\label{S-apdx}

\begin{figure}
\begin{center}
\includegraphics[width=\textwidth,height=0.5\textheight,clip]{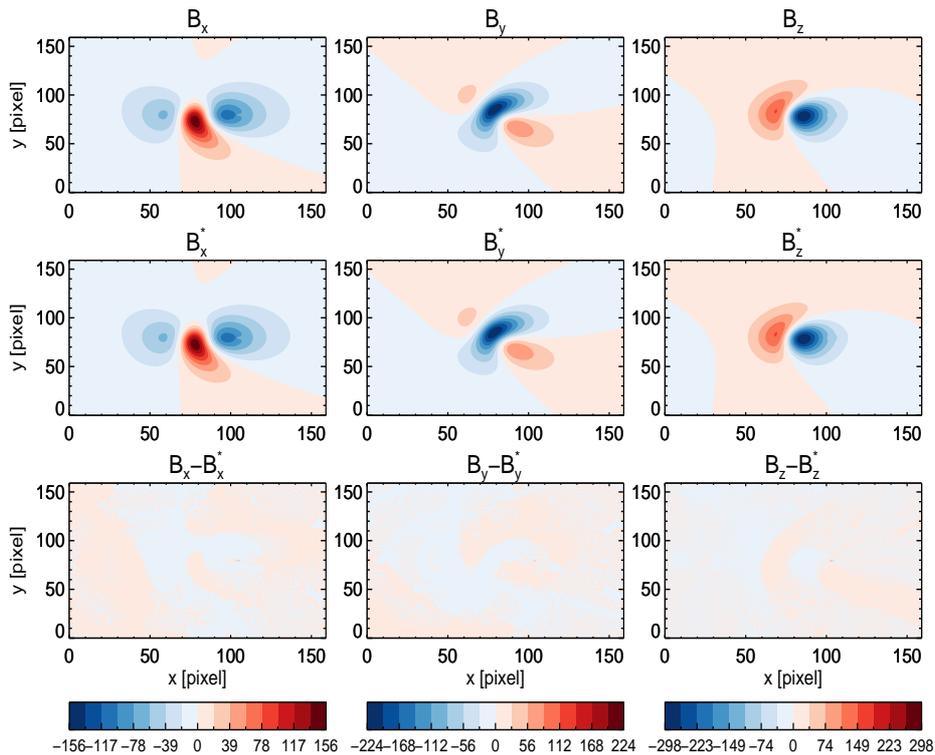}
\caption{Comparison of the Low and Lou magnetic field $\mathbf{B}$ (first row) with the reconstructed one $\mathbf{B}^*=\nabla\times\mathbf{A}$ (second row) at the bottom boundary $(x,y,z=0)$. The third row shows the difference $\mathbf{B}-\mathbf{B}^*$. The left, middle, and right columns show the $x$, $y$, and $z$ components of the magnetic fields respectively.}
\label{figll}
\end{center}
\end{figure}

Here we test our potential-field and vector-potential methods  against the well-known analytical nonlinear force-free fields of \inlinecite{lowlou90}. More specifically, we compare with the second \citeauthor{lowlou90} case, namely the one with $n=m=1$ (in their notation) and source location parameters $l=0.3$, $\Phi=\pi/4$. For this we use a rectangular box of size $2\times2\times1.6$ (in arbitrary units) that is converted into a grid of $160\times160\times128$ pixels. After calculating the vector potential $\mathbf{A}$ from Equations~(\ref{vpot})-(\ref{vpot4}) we reconstruct the field $\mathbf{B}$ as $\mathbf{B}^*=\nabla\times\mathbf{A}$. A visual comparison between the original and reconstructed fields at the lower boundary of the volume is shown in Figure~\ref{figll}. We notice that the method reproduces the given field quite well.

It is necessary, nonetheless, to quantify the agreement between fields $\mathbf{B}$ and $\mathbf{B}^*$. To this purpose we utilize the linear correlation coefficients for each component of the two fields, as well as the metrics described by \inlinecite{schrijver06}. Starting with the latter, the vector correlation of $\mathbf{B}$, $\mathbf{B}^*$ is $C_{\rm vec}=0.9967$, the Cauchy-Schwarz metric is $C_{\rm CS}=0.9984$, the complement of normalized vector error is $E_{\rm n}'=0.9747$, the complement of mean vector error is $E_{\rm m}'=0.9496$ and the total magnetic energy normalized to the input case is $\epsilon=1.013$. The correlation coefficients for the $x$ and $y$ components are practically 1, while the one for $B_z$ is slightly lower, $\sim 0.9935$. This is a general rule, with the $z$-component of the field reproduced less well than the $x$ and $y$-components, owning to the setup of the method. Indeed, with the used gauge the calculation of $B_z$ involves the most numerical operations, therefore it is susceptible to more accumulated errors. The corresponding metrics for the potential fields $\mathbf{B}_\mathrm{p}$, $\mathbf{B}_\mathrm{p}^*=\nabla\times\mathbf{A}_\mathrm{p}$ have similar values, indicating a good agreement for these fields as well.

Another remark is that if we take as reference the top plane (at $z=z_2$) in the calculation of the vector potentials, the agreement between the two fields is a bit better. As an example, the vector correlation of $\mathbf{B}$, $\mathbf{B}^*$ for the top level is $C_{\rm vec}=0.9997$ and the complement of mean vector error is $E_{\rm m}'=0.9951$. Improved precision in case the top plane is used as reference seems a general characteristic of the method seen in other examples, as well.

If we calculate the total relative helicity from Equation~(\ref{tothel}) using the volume-calculation method we obtain $H=-0.495\,H_\mathrm{LL}$, where $H_\mathrm{LL}=|\int_\mathcal{V} \mathrm{d}V\,\mathbf{A}_\mathrm{LL}\cdot\mathbf{B}|$ is a non-invariant helicity. $\mathbf{A}_\mathrm{LL}$ is the vector potential that produces the \citeauthor{lowlou90} field, given by
\begin{equation}
\mathbf{A}_\mathrm{LL}=\frac{a}{r^2}\int_{-1}^\mu d\mu'\frac{P^2(\mu')}{1-\mu'^2}\hat{r}+\frac{P(\mu)}{r^2\sin\theta}\hat{\phi},
\end{equation}
where $a^2=0.425$ and $P=P_{1,1}$ in \citeauthor{lowlou90}'s notation. One can easily verify that $\mathbf{B}=\nabla\times\mathbf{A}_\mathrm{LL}$. If we replace $\mathbf{A}$ in Equation~(\ref{tothel}) with $\mathbf{A}_\mathrm{LL}$ we obtain $H=-0.482\,H_\mathrm{LL}$, so our implementation of the \inlinecite{val12} method reproduces the actual helicity of the \citeauthor{lowlou90} model within $\sim2.7\%$ of accuracy.

The energy of the field, as calculated by Equation~(\ref{ecdef}), is $E_\mathrm{t}=45.0$ in the arbitrary units used so far in this section. Similarly, the potential energy is $E_\mathrm{p}=34.8$, and so the free magnetic energy is $E_\mathrm{c}=0.226\,E_\mathrm{t}$, a fraction that agrees with the one found by \citeauthor{lowlou90}, although for a slightly different case ($\Phi=\pi/2$).

\begin{acks}
This research has been carried out in the framework of the research projects  hosted by the RCAAM of the Academy of Athens. The observations are used courtesy of NASA/SDO and the HMI science team. We thank X.~Sun and Y.~Liu for the provided magnetic field extrapolations of NOAA AR 11158. The simulations were performed on the STFC and SRIF funded UKMHD cluster, at the University of St Andrews. VA acknowledges support by EU (IEF-272549 grant). This work was supported from the EU's Seventh Framework Programme under grant
agreement n$^o$ PIRG07-GA-2010-268245. It has been also co-financed by the European Union (European Social Fund -- ESF) and Greek national funds through the Operational Program "Education and Lifelong Learning" of the National Strategic Reference Framework (NSRF) -- Research Funding Program: Thales. Investing in knowledge society through the European Social Fund.
\end{acks}

\bibliographystyle{spr-mp-sola}
\bibliography{revised_ms}

\end{article}
\end{document}